\newcommand{\ta}[1]{\left[ #1 \right]_t}

\newcommand{\U}{\uparrow}
\newcommand{\D}{\downarrow}

\documentclass[%
reprint,amsmath,amssymb,aps,prl,superscriptaddress,longbibliography,
]{revtex4-1}

\usepackage{graphicx}
\usepackage{graphics}
\usepackage{dcolumn}
\usepackage{bm}
\usepackage{epstopdf}
\usepackage{float}
\usepackage{tikz}
\usepackage{MnSymbol}
\usepackage{wasysym}
\usepackage{lipsum}
\usepackage{braket}
\usepackage{cellspace} %
\usepackage{makecell} %
\begin{document}
\title{Adiabatic quantum state transfer in a semiconductor quantum-dot spin chain}

\author{Yadav P. Kandel}
%
\author{Haifeng Qiao}
%
\affiliation{Department of Physics and Astronomy, University of Rochester, Rochester, NY, 14627 USA}
\author{Saeed Fallahi}
\affiliation{Department of Physics and Astronomy, Purdue University, West Lafayette, IN, 47907 USA}
\affiliation{Birck Nanotechnology Center, Purdue University, West Lafayette, IN, 47907 USA}
\author{Geoffrey C. Gardner}
\affiliation{Birck Nanotechnology Center, Purdue University, West Lafayette, IN, 47907 USA}
\affiliation{School of Materials Engineering, Purdue University, West Lafayette, IN, 47907 USA}
\author{Michael J. Manfra}
\affiliation{Department of Physics and Astronomy, Purdue University, West Lafayette, IN, 47907 USA}
\affiliation{Birck Nanotechnology Center, Purdue University, West Lafayette, IN, 47907 USA}
\affiliation{School of Materials Engineering, Purdue University, West Lafayette, IN, 47907 USA}
\affiliation{School of Electrical and Computer Engineering, Purdue University, West Lafayette, IN, 47907 USA}
\author{John M. Nichol}
\affiliation{Department of Physics and Astronomy, University of Rochester, Rochester, NY, 14627 USA}
\affiliation{Corresponding author: john.nichol@rochester.edu}

\begin{abstract}
\textbf{Abstract} Semiconductor quantum-dot spin qubits are a promising platform for quantum computation, because they are scalable and possess long coherence times. In order to realize this full potential, however, high-fidelity information transfer mechanisms are required for quantum error correction and efficient algorithms. Here, we present evidence of adiabatic quantum-state transfer in a chain of semiconductor quantum-dot electron spins. By adiabatically modifying exchange couplings, we transfer single- and two-spin states between distant electrons in less than 127 ns. We also show that this method can be cascaded for spin-state transfer in long spin chains. Based on simulations, we estimate that the probability to correctly transfer single-spin eigenstates and two-spin singlet states can exceed 0.95 for the experimental parameters studied here. In the future, state and process tomography will be required to verify the transfer of arbitrary single qubit states with a fidelity exceeding the classical bound. Adiabatic quantum-state transfer is robust to noise and pulse-timing errors. This method will be useful for initialization, state distribution, and readout in large spin-qubit arrays for gate-based quantum computing. It also opens up the possibility of universal adiabatic quantum computing in semiconductor quantum-dot spin qubits.
\end{abstract}
\maketitle
\section{Introduction}

Progress towards fabrication of large spin-qubit arrays~\cite{Zajac2016,Volk2019}, together with methods for orthogonal control of quantum-dot chemical potentials~\cite{Baart2016single,Mills2019CS,Volk2019}, inter-dot tunnel couplings~\cite{Hensgens2017,VanDiepen2018,Mills2019CA,Hsiao2020,zwolak2020autotuning}, and nearest neighbor exchange couplings~\cite{qiao2020}, have opened up the possibilities of implementing complex multi-qubit quantum operations~\cite{Kandel2019,qiao2020conditional} in semiconductor quantum-dot spin qubits. To tap the full potential of these developments, and to realize a large scale fault-tolerant quantum computer, high-fidelity information transfer mechanisms between qubits are required. Since quantum-dot spin qubits naturally interact through the nearest-neighbor Heisenberg exchange coupling, long distance inter-qubit coupling is challenging. Quantum information transfer has been achieved in spin-qubits by electron shuttling using electrical pulses~\cite{Baart2016single,Fujita2017,Flentje2017coherent,Nakajima2018}, mechanical waves~\cite{bertrand2016fast}, spin SWAP operations~\cite{Kandel2019,Sigillito2019}, and quantum mediators~\cite{Baart2017,Malinowski2019}. These methods, elegant as they are, have their limitations, often including stringent pulse-timing requirements. In this work, we report evidence for the successful experimental implementation of adiabatic evolution methods to achieve quantum information transfer in a chain of quadruple quantum-dots. Compared to conventional pulsed information transfer methods, adiabatic techniques are more robust to pulse errors and system noise.

Adiabatic quantum information processing in arrays of spin qubits has been the focus of intense theoretical research~\cite{farhi2000,Bacon2009AGT,Lloyd2004,Srinivasa2007,Srinivasa2009,Oh2013,Menchon2016,Platero2019,Petrosyan2010,Chancellor2012,Farooq2015}, due to the possibility of high-fidelity operations in the presence of noise or pulse errors. Adiabatic shuttling of spin states has been already demonstrated via electron shuttling~\cite{Baart2016single,Fujita2017,Flentje2017coherent,Nakajima2018}. Here, we present evidence for adiabatic quantum-state transfer (AQT) of both single-spin eigentsates and two-spin singlet states in a GaAs quadruple quantum-dot device. Unlike previous works, this approach does not involve the physical motion of electrons. Specifically, we design a time-dependent Hamiltonian for a linear chain of three electron spins. As the spins evolve under the action of the Hamiltonian, an initial state of the first spin is transferred to the third spin. This process is closely related to stimulated adiabatic Raman passage, a time-honored technique from the optical physics community~\cite{Vitanov2017}, which has been implemented in other qubit platforms~\cite{Vitanov2017,Kumar2016apsc,wunderlich2007robust,Vepsalainen2019superad}. Also, the process we use is identical to adiabatic quantum teleportation~\cite{Oh2013,Bacon2009AGT}. We show that the AQT process can be cascaded to transfer spin states across a longer spin chain.

We simulate our experiments, taking into account known sources of errors and noise~\cite{supmat}, and we find that the results of our simulations closely match the experimental data. Based on those simulations, we estimate that the probability to correctly transfer a single-spin eigenstate or a two-spin singlet-state can exceed 0.95, in operation times of less than 127 ns. In lieu of full quantum-state tomography, which would require a micromagnet~\cite{Pioro-Ladriere2008} or an antenna~\cite{Koppens2006} for magnetic resonance, we implement different quantum gates to assess the spin states after AQT. In the future, state and process tomography will be required to verify the AQT performance. The main limiting factor of the AQT fidelity in our experiment is the nuclear hyperfine noise in the GaAs/AlGaAs heterostructure, which gives rise to a fluctuating magnetic-field gradient between dots. In Si devices, we expect that high-fidelity transfer of arbitrary single-qubit states could easily be achieved~\cite{Oh2013,gullans2020SCTAP,supmat}.

\section{Results}
\subsection{Device}
\begin{figure}
	\includegraphics{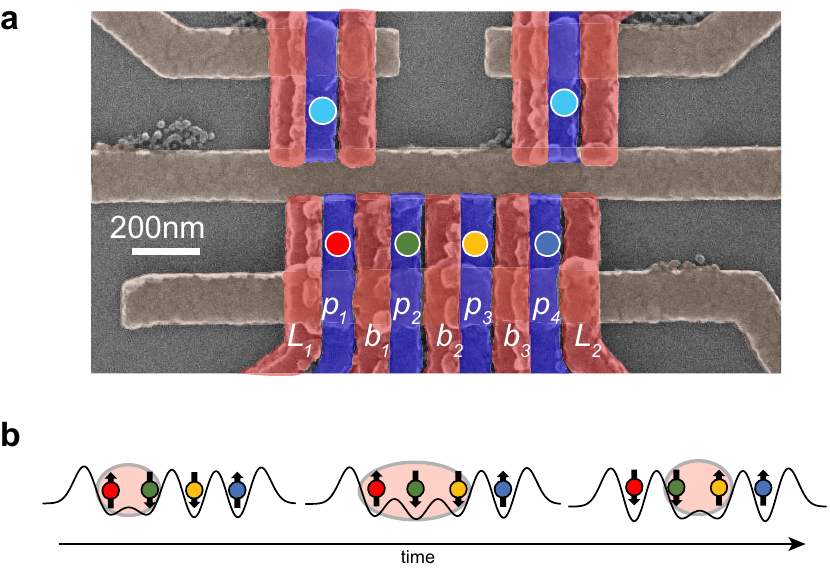} 
	\caption{\label{fig1} Experimental setup. (a) False-color scanning electron micrograph of a quadruple quantum dot device similar to the one used in the experiment. The quantum dots are located in 2DEG below the positions marked by circles. Voltages applied to three layers of metal gates (brown, red, and blue) create the quantum-dot confinement potentials. A top gate, which is not present in this figure, covers the active area of the device. (b) Schematic showing the changes in the quantum-dot barrier heights for an AQT process that transfers the state of qubit $3$ to qubit $1$. This process also transfers the state of qubits 1-2, which are in an eigenstate of exchange, to 2-3.}
\end{figure}

Our quadruple-quantum-dot device with overlapping gates is fabricated in a GaAs/AlGaAs heterostructure [Fig. 1(a)]~\cite{Angus2007,Zajac2016}. Two additional dots above the main array are configured for readout via rf-reflectometry~\cite{Barthel2010}. We divide the quadruple quantum dot array into two pairs for initialization and measurement. Dots 1 and 2 form the ``left" side and dots 3 and 4 form the ``right" side. We measure the left and right pairs in the two-electron singlet/triplet basis using Pauli spin blockade together with a shelving mechanism ~\cite{Studenikin2012}. The singlet is $\ket{S}=\frac{1}{\sqrt{2}}\left(\ket{\U\D}-\ket{\D\U}\right)$, and the triplets are  $\ket{T^0}=\frac{1}{\sqrt{2}}\left(\ket{\U\D}+\ket{\D\U}\right)$, $ \ket{T^+}=\ket{\U\U}$, and $\ket{T^-}=\ket{\D\D}$.  The device is operated at the symmetric tuning~\cite{Reed2016SymSi,Martins2016}, where each dot contains one electron, and all chemical potentials are roughly the same. We independently control the exchange couplings between dots using the techniques described in Ref.~\cite{qiao2020}. The state preparation and readout is further described in Methods.

The linear chain of spins with time-dependent nearest-neighbor exchange coupling in our device can be modeled using the Heisenberg model and the Hamiltonian is
\begin{equation}
H(t)  = \frac{h}{4}  \sum_{i=1}^{3} J_{i}\left(t\right) \bm{ \sigma}_i\cdot \bm{\sigma}_{i+1} + \frac{h}{2}\sum_{i=1}^{4} B_i^z\sigma_i^z, \label{eq:1}
\end{equation}
where $J_{i}(t)$ is nearest-neighbor exchange interaction between spins in quantum dots $i$ and $i+1$ at time $t$, and $B_i^z$ is the $z-$component of the magnetic field at the location of dot $i$. Both $J_i$ and $B^z_i$ have units of frequency. $\bm{\sigma}_i =[\, \sigma_i^x, \sigma_i^y, \sigma_i^z]\,$ is the Pauli vector operating on spin $i$, and $h$ is the Planck constant. $B_i^z$ accounts for both the external field of $0.5$ T applied to polarize the spin states and the local hyperfine field~\cite{supmat}. Because the $x-$ and $y-$ components of the hyperfine field are negligible compared to external magnetic field, they are omitted in the second term of this Hamiltonian.

\subsection{Adiabatic quantum-state transfer}
\begin{figure}
\includegraphics{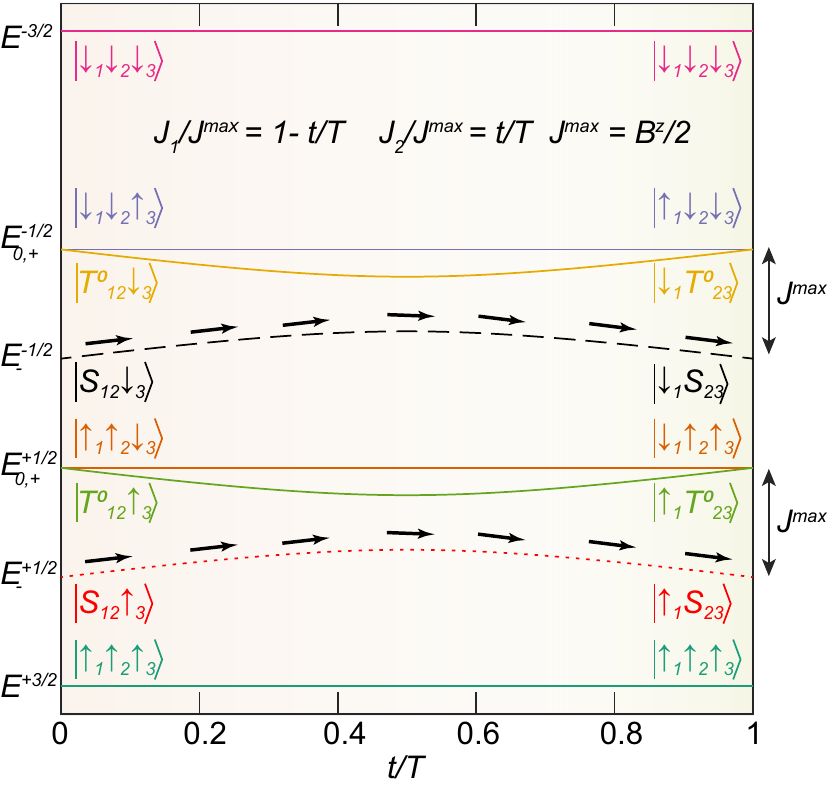} 
\caption{\label{fig2} Eigenstates of the time-dependent three-spin Heisenberg Hamiltonian with $[J_1(t),\ J_2(t)]=J^{max}[1-t/T,\ t/T]$ for $0<t<T$. $B^z$ is the uniform magnetic field. The eigenstates at the initial and final times are labeled. Adiabatic state transfer can occur by initializing the system in either the $E_-^{+1/2}$ or $E_-^{-1/2}$ states. Here, the superscript is $z-$component of the spin angular momentum $(S^z)$, and subscripts denote different eigenstates within a particular $S^z$ subspace.}
\end{figure}

\begin{figure*}
\includegraphics{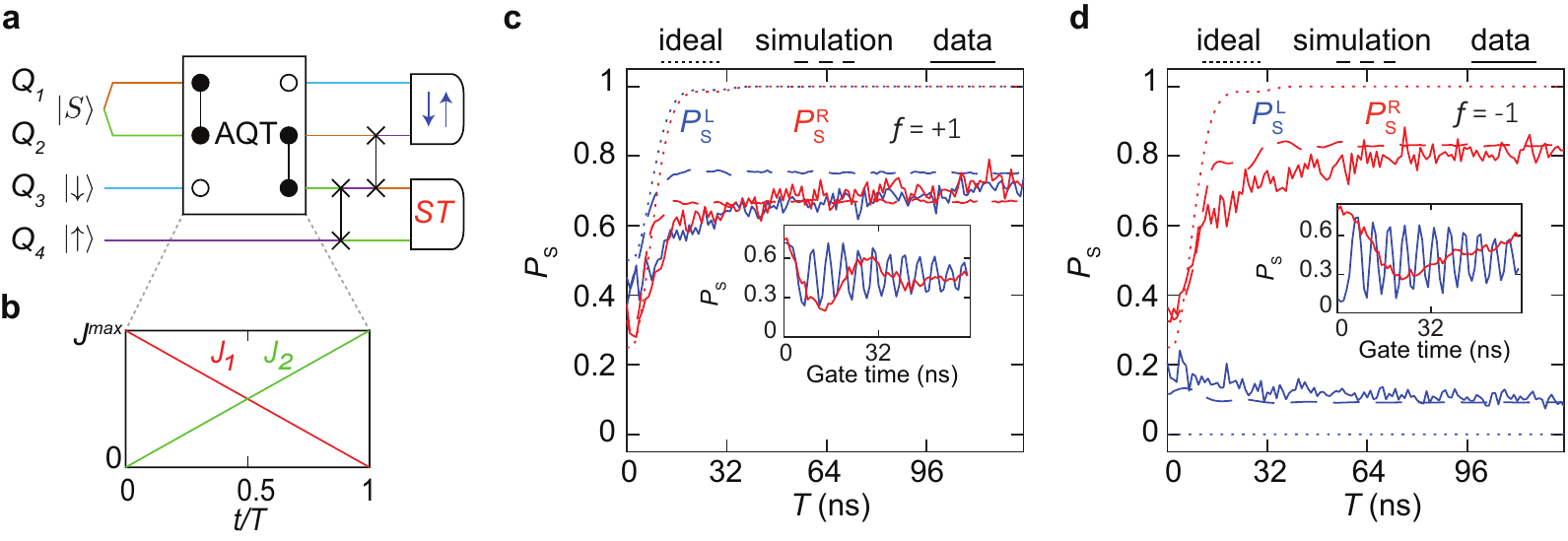}
\caption{\label{fig3} Three-spin AQT. (a) Quantum circuit diagram for the experiment. The spin chain is initialized as $\ket{S_{12}\D_3\U_4}$ and the AQT implemented in spins 1-3 transfers the state of spin 3 to spin 1 and the singlet state in spins 1-2 to spins 2-3. Then, spins 3-4 and 2-3 are swapped, in this order. We then measure the left pair $(P_\text{S}^\text{L})$ and the right pair $(P_\text{S}^\text{R})$ in the singlet/triplet basis via Pauli spin blockade. The colors represent the physical locations of the initial states. (b) Change in exchange coupling strengths between qubits for the AQT step in (a). Here, $T$ is the Hamiltonian interpolation time and $0<t<T$. (c) Singlet return probabilities of the left and right pairs as a function of interpolation time $T$ for $f=+1$. (d) Same as (c), but for $f=-1$. In both (c) and (d), the expected outcomes under ideal conditions (dotted lines) as well as simulated results including known errors and noise (dashed lines) are overlaid on top of the measured data (solid lines). The insets in (c) and (d) show exchange oscillations in spins 1-2 and $S-T^0$ oscillation in spins 3-4 after the experiment described in (a). ``Gate time" refers to these oscillation times. The presence of exchange oscillations in spins 1-2 and $S-T^0$ oscillations in spins 3-4 provides evidence of the successful adiabatic transfer. Each data point represents the average of 512 single-shot measurements.}
\end{figure*}

To implement adiabatic spin-state transfer, we initialize the spin chain in the state $\ket{S_{12}\D_3\U_4}$ or $\ket{S_{12}\U_3\D_4}$.  Dots $3$ and $4$ contain spins in the $\ket{\D_3\U_4}$ or $\ket{\U_3\D_4}$ configuration depending on the sign of the hyperfine gradient associated with dots $3$ and $4$~\cite{Petta2005,foletti2009universal}. 

Once the spin chain is initialized, we set $[ J_{1}(t),\ J_{2}(t),\ J_{3}(t)] = J^{max} [ 1-t/T,\ t/T,\ 0 ]$ for $0<t<T$, with $J^{max}=120$ MHz. Note that the initial state discussed above is an eigenstate of $H(0)$ when $J_1(0)\gg \left|B_2^z-B_1^z\right|$. Figure~\ref{fig2} shows the time-dependent eigenvalues of the three-spin analog of this Hamiltonian for a related configuration of exchange couplings. When $H(t)$ changes adiabatically, and because $H(t)$ conserves both the total angular momentum and the $z$-component of angular momentum, $S^z$, a particular eigenstate at $t=0$ is mapped to an eigenstate with the same $S^z$ at time $t=T$. In particular, an initial state of spins 1-3, $\ket{S_{12}\D_3}$, transitions to $\ket{\D_1 S_{23}}$, as shown in Fig.~\ref{fig2}. Likewise, $\ket{S_{12}\U_3}$ transitions to $\ket{\U_1 S_{23}}$. One can view this process as transferring the state of dot 3 to dot 1, while simultaneously transferring the joint spin state of dots 1-2, which is a singlet state, to dots 2-3. In principle, AQT can transfer an arbitrary spin state of dot 3 to dot 1. As we discuss further below, AQT can transfer two-spin states in addition to singlets, although it generally performs best if spins 1 and 2 are configured as a singlet. Figure~\ref{fig1}(b) illustrates the physical implementation of this AQT process.

To measure the spin states after the AQT process, we apply SWAP operations~\cite{Kandel2019} between spins 3-4 and 2-3, in this order, to bring the singlet state to the right pair and the product state to the left pair of spins before measurement [Fig.~\ref{fig3}(a)]. We measure the left pair by adiabatic projection and the right pair by diabatic projection onto the singlet/triplet basis~\cite{Petta2005,foletti2009universal}. Diabatic projection preserves the singlet state, and adiabatic projection maps either $\ket{\U\D}$ or $\ket{\D\U}$ to the singlet, and all other states to the triplets, depending on the sign of the hyperfine gradient~\cite{Petta2005,foletti2009universal,supmat}.

\subsection{Effects of the nuclear hyperfine gradient}
Since the initial product state of the left pair is eventually measured in the right pair, knowledge of the magnetic-field gradients in both pairs is required for proper interpretation of the experimental data. We define $f = \text{sign}(B_2^z-B_1^z)\times\text{sign}(B_4^z-B_3^z)$. When $f=+1$, both pairs have gradients of the same sign, and when $f=-1$, the pairs have gradients with opposite signs. To measure $f$, we initialize both sides as product states with $S^z=0$. Then, we evolve spins 2-3 under exchange coupling for variable amount of time. When $f=+1$, corresponding to initial states $\ket{\U_1\D_2\U_3\D_4}$ or $\ket{\D_1\U_2\D_3\U_4}$, prominent exchange oscillations are visible. When $f=-1$, corresponding to initial states $\ket{\D_1\U_2\U_3\D_4}$ or $\ket{\U_1\D_2\D_3\U_4}$, spins 2 and 3 have the same orientation, and no exchange oscillations occur. We interleaved these measurements of $f$ with measurements of the AQT process to distinguish the $f=\pm 1$ cases. 

Figures~\ref{fig3}(c)-(d) show the results of the experiment described by the circuit of Fig.~\ref{fig3}(a) for the $f=+1$ and $f=-1$ cases, respectively. Calculated outcomes for the ideal cases, and simulation results taking into account all known sources of noise and errors, are overlaid on top of the data. The simulation results match the measurements in both cases~\cite{supmat}. In Figs.~\ref{fig3}(c)-(d), the gradual rise in the return probability with $T$ occurs because for small values of $T$, the process is not sufficiently adiabatic. At large values of $T$, the return probabilities saturate, suggesting successful adiabatic transfer. The predicted oscillations in the return probability at small values of $T$ are related to resonant adiabatic transfer, which we discuss further below.
 
Each data point in Fig.~\ref{fig3} is averaged over 512 single-shot measurements for each value of $T$. We repeat this sequence of 512 single-shot measurements 256 different times. Each repetition takes no more than one second to acquire, and the hyperfine gradients and the value of $f$ are empirically quasi-static during each repetition. Different repetitions were thus used for the $f=+1$ and $f=-1$ cases shown in Fig.~\ref{fig3}. The full dataset, which includes all repetitions, is shown in Methods. We display single repetitions here, because the approximately constant value of the hyperfine field during a single repetition enables accurate simulation. 

As discussed further in Methods, the sign of $f$ changes on a timescale ranging from seconds to tens of seconds, and typical gradient strengths are on the order tens of MHz. These values are consistent with previous estimates of nuclear spin diffusion times and rates in GaAs double quantum dots~\cite{Reilly2010,Shulman2014}.

\begin{figure*}
	\includegraphics{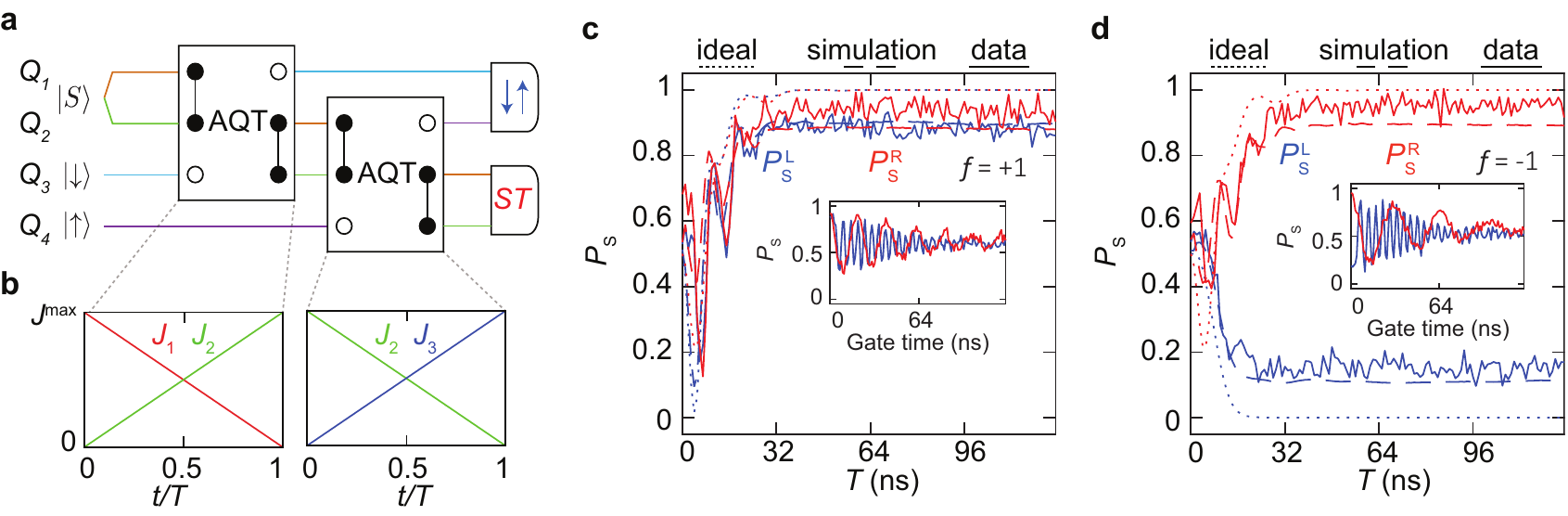}
	\caption{\label{fig4} AQT Cascade. (a) Quantum circuit diagram for the experiment. We initialize the spins as $\ket{S_{12}\D_3\U_4}$. Applying two AQT steps transfers the product state of spins 3-4 to 1-2, and the singlet state of spins 1-2 to spins 3-4. (b) Exchange coupling parameters as a function of time for the AQT steps shown in (a), where $T$ is the Hamiltonian interpolation time and $0<t<T$. (c) Singlet return probabilities for the left pair $(P_\text{S}^\text{L})$  and the right pair $(P_\text{S}^\text{R})$  when $f=+1$. (d) Data for $f=-1$. Simulations neglecting and including known sources of error are overlaid in each panel. The insets in (c) and (d) show prominent exchange oscillations between spins 1-2 and singlet-triplet oscillations associated with spins 3-4, after the AQT cascade described in (a), providing evidence of successful transfer of spin states. Each data point represents the average of 256 single-shot measurements.}
\end{figure*}

\subsection{Coherent evolution after AQT}
To further assess the AQT, we use additional quantum gates to test the spin states. First, we transfer the spin states as described above with $J^{max}=120$ MHz and $T=127$ ns. Then, we perform the SWAP gates discussed previously. In the case of successful state transfer, the initial product state of spins 3-4 occupies spins 1-2, and the initial singlet state of spins 1-2 occupies spins 3-4. Then, we induce exchange coupling $J$ between spins 1 and 2 for a variable amount of time. The measured singlet-return probability of spins 1 and 2 will contain an oscillatory component of the form $P_S(t)=A\cos(2 \pi J t+\theta)$, when the joint state $\ket{\psi}$ of those spins has a component of the form $\sqrt{2 A} \left( \cos{(\theta/2)} \ket{\U \D}+i\sin{(\theta/2)}\ket{\D \U} \right)$, when $B^z_1 > B^z_2$, and where $0 \leq A \leq 1/2$. In the language of singlet-triplet qubits, exchange oscillations will occur if the joint spin-state has a component in the $x-y$ plane of the Bloch sphere, where the $z$ axis is defined by the exchange coupling. When $A \approx 1/2$ and $\theta \approx 0$, we infer that $\ket{\psi}$ has a large component along $\ket{\U \D}$. When $\theta \approx \pi$, we infer that  $\ket{\psi}$ has a large component along $\ket{\D \U}$.
	
We also allow spins 3 and 4 to evolve for a variable length of time in the presence of a hyperfine gradient while separated. In this case, the measured singlet-return probability of spins 3 and 4 will contain an oscillatory component of the form $P_S(t)=A\cos(2 \pi (B_4^z-B_3^z)t+\theta)$, when their joint state $\ket{\psi}$ has some overlap with a state of the form $\sqrt{2 A} \left(\cos{(\theta/2)} \ket{S}+i\sin{(\theta/2)}\ket{T^0} \right)$, where $0 \leq A \leq 1/2$. In the language of singlet-triplet qubits, singlet-triplet oscillations will occur if the joint spin-state has a component in the $y-z$ plane of the Bloch sphere, where the $x$ axis is defined by the hyperfine gradient. When $A \approx 1/2$ and $\theta \approx 0$, we infer that $\ket{\psi}$ has a large component along $\ket{S}$.

The results of these experiments are shown in the insets of Figs.~\ref{fig3}(c)-(d)~\cite{supmat}. The presence of large-amplitude oscillations on both sides with the expected phases provides further evidence of successful transfer of both spin-up and spin-down eigenstates from spin 3 to 1 and a singlet state from spins 1-2 to 2-3 during the AQT process. Note that the exchange oscillations of spins 1-2 have different phases for $f = \pm 1$, as expected. 

Our data provide strong evidence that both single-spin eigenstates and two-spin singlet states, which are also eigenstates of the exchange operator, are correctly transferred by the AQT process. Figs.~\ref{fig3}(c) and (d) provide evidence that spin eigenstates can be transferred from dot 3 to dot 1. The transfer of the singlet state can be viewed as the corresponding process that transfers the state of dots 1-2 to 2-3. The coherent evolution of the singlet state after the AQT process provides evidence of its successful transfer. In the future, as discussed further below, complete state and process tomography will be required to assess the performance of AQT for arbitrary quantum states.

\subsection{Relationship to counterintuitive adiabatic transfer} 
The AQT sequence described above, which transfers the state of dot 3 to dot 1, partially resembles the ``counterintuitive" adiabatic transfer sequence used in optical systems~\cite{Vitanov2017, Oh2013}.  The sequence of Fig.~\ref{fig3}(a) transfers a spin state from dot 3 to dot 1, yet the sequence begins with a strong exchange coupling between dots 1 and 2, neither of which contain the state to be transferred. 

However, a true counterintuitive adiabatic process relies on the existence of a ``dark state,'' which contains no excitation of the intermediate state. In the present case, the desired dark state would feature no evolution of the second spin in a three-spin chain. Although it is possible to create a dark state in a spin chain with an $XX$ (Ising) coupling~\cite{Oh2013}, the dark state does not occur for the general case of an exchange-coupled  (Heisenberg) spin chain~\cite{Oh2013}. However, specific combinations of the exchange-couplings and magnetic-field differences between dots can yield a true counterintuitive adiabatic sequence~\cite{Oh2013,gullans2020SCTAP}.

In general, it is possible to implement the AQT process described above with spins 1 and 2 configured as any eigenstate of the exchange operator, including either of the polarized triplet states, which do not evolve under exchange coupling. However, as Fig.~\ref{fig2} shows, many of the configurations involving other eigenstates of exchange pass through degeneracies at the beginning and the end of the time evolution, complicating the transfer process. We return to this point below when we estimate the fidelity of the AQT process.

\subsection{AQT Cascade}
The AQT process described above transfers spin states between three electrons. We now show that AQT processes can be cascaded to enable long-distance state transfer. We use two AQT steps in a chain of four spins [Fig.~\ref{fig4}(a)]. We initialize the spin chain in the state $\ket{S_{12}\D_3\U_4}$ (or $\ket{S_{12}\U_3\D_4})$. In the first AQT, we set $[J_{1}(t),\ J_{2}(t),\ J_{3}(t) ]= J^{max} [1-t/T,\ t/T,\ 0]$ for $0<t<T$, where $J^{max}= 120$ MHz and $T$ ranges from 0 to 127 ns [Fig.~\ref{fig4}(b)].  In the adiabatic limit, the spin state from spin 3 transfers to spin 1, and the singlet state in spins 1-2 transfers to spins 2-3 so that the state of the spin chain becomes $\ket{\D_1S_{23}\U_4}$ (or $\ket{\U_1S_{23}\D_4}$). In the second AQT step, we set $[J_{1}(t), J_{2}(t), J_{3}(t)]= J^{max}[0,\ 1-t/T,\ t/T]$ for $0<t<T$. In the adiabatic limit, this process transfers the spin state of spin $4$ to spin $2$, and the singlet state in spins 2-3 transfers to spins 3-4 so that the final state of the spin chain becomes $\ket{\D_1\U_2S_{34}}$ (or $\ket{\U_1\D_2S_{34}}$). We measure the left and right pairs as before. 

Figures~\ref{fig4}(c)-(d) show the cases for $f=+1$ and $-1$, respectively. Even though the data of Figs.~\ref{fig4}(c)-(d) involve two AQT steps, the maximum transfer probability appears higher than the data of Figs.~\ref{fig3}(c)-(d), which involve one AQT step and two SWAP gates. We attribute this difference to the relative insensitivity of the AQT process to noise and pulse errors, as compared to the SWAP gates. This difference highlights the robustness and potential usefulness of AQT in quantum-dot spin chains. Our simulations agree with our measurements.

As before, we induce exchange between spins 1-2 and singlet-triplet evolution between spins 3-4 following the state transfer~\cite{supmat}. The data from these measurements are shown in the insets of Figs.~\ref{fig4}(c)-(d). The presence of prominent oscillations with the expected phases in both cases provides further evidence of successful transfer of single-spin eigenstates and two-spin singlet states.

As in Figs.~\ref{fig3}(c)-(d), the data of Figs.~\ref{fig4}(c)-(d) show oscillatory features at small values of $T$, which are related to resonant adiabatic quantum-state transfer~\cite{Oh2013}. These resonances in the non-adiabatic limit provide a shortcut to adiabatic quantum-state transfer. In the present experiment, effects associated with the hyperfine gradient broaden and reduce the overall height of the resonant peaks. We expect that resonant adiabatic transfer should work better in Si spin qubits, where nuclear spin effects are suppressed.

\begin{figure}
\includegraphics{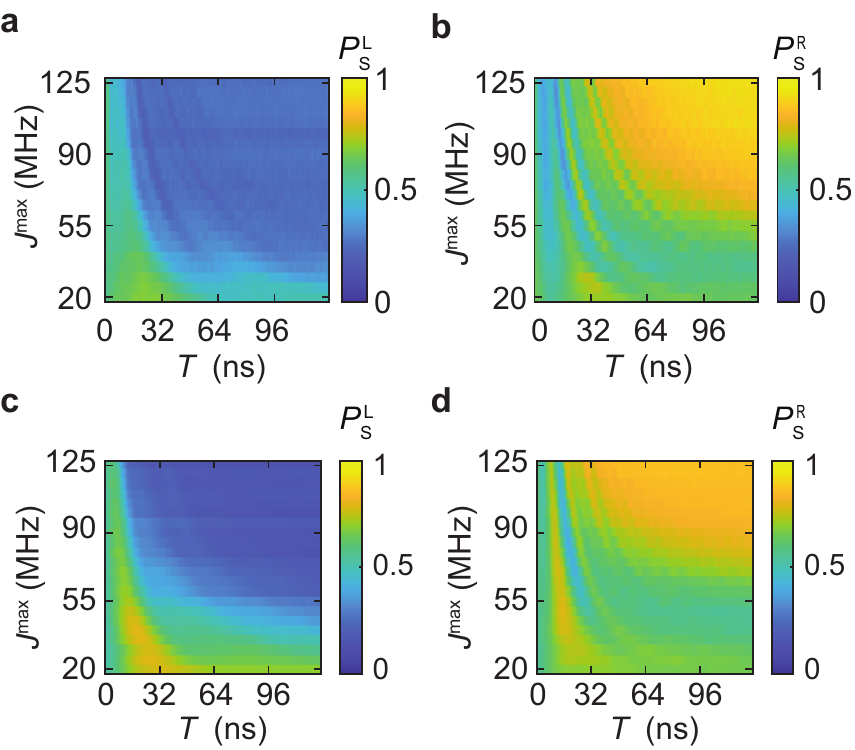} 
\caption{\label{fig5} Effects of maximum coupling strength $J^{max}$ and interpolation time $T$. (a) Singlet return probability of the left pair, and (b) the right pair as a function of $ J^{max} $ and $T$. (c-d) Simulations corresponding to (a) and (b), respectively. During this experiment, we observed that in most cases $f=-1$. Data presented in the first row are post-selected for $f=-1$ ~\cite{supmat}.}
\end{figure}

To further explore effects associated with the speed of the state transfer, we plot measurements of the cascaded AQT probability as we vary $T$ and $J^{max}$ (Fig.~\ref{fig5}). We find that increasing $J^{max}$ or $T$ both correlate with higher transfer probability. This is expected, because the condition for adiabatic transfer is $J^{max}T/\hbar>>1$~\cite{Oh2013}. We also observe prominent features associated with resonant adiabatic transfer, especially at low values of $T$. Although harnessing resonant adiabatic transfer requires more precise control pulses than adiabatic transfer, it provides a route to distant state transfer in shorter times than adiabatic transfer.

\subsection{Fidelity Estimate}
Although we have not performed state tomography to definitively assess the AQT performance, we may estimate its fidelity as follows. The simulations presented in Figs.~\ref{fig3}, \ref{fig4}, and \ref{fig5} take into account state preparation and measurement (SPAM) errors, as well as hyperfine noise, low-frequency charge noise, high-frequency charge noise, and pulse imperfections. The levels of average hyperfine fields and their fluctuations are chosen to replicate the observed coherence of singlet-triplet oscillations in our device. The magnitude of the low-frequency exchange coupling noise is chosen to replicate the observed exchange oscillation quality factors in our device. We have also included white high-frequency charge noise~\cite{supmat}. The simulations show good quantitative agreement with our data. 

To estimate the probability of correctly transferring a single-spin eigenstate via a single AQT process, we simulate a three spin system in the initial state $\ket{\psi_0}=\ket{S_{12} \phi_3}$, where $\ket{\phi}$ is a single-spin eigenstate. We numerically evolve this state in time under the AQT Hamiltonian with $J^{max}=120$ MHz~\cite{supmat} to a final state $\ket{\psi_f}$. We include noise and pulse errors in this simulation. We neglect SPAM errors in this calculation to assess the performance of the AQT operation itself. For each instance of the simulation, we trace out spins 2 and 3 from the final state to obtain a reduced density matrix for spin 1: $\rho_{f,1}$. Setting $\rho_{i,1}=\ket{\phi}\bra{\phi}$, we compute the state fidelity as $f_1=\left[\textnormal{Tr}\left(\sqrt{\sqrt{\rho_{f,1}}\rho_{i,1}\sqrt{\rho_{f,1}}}\right)\right]^2$. To estimate a transfer probability for single-spin eigenstates, we average the resulting state fidelity over different charge-noise and hyperfine-noise configurations and over initial states in the set $\ket{\phi}=\{\ket{\U},\ket{\D}\}$. To establish a transfer fidelity for the singlet state, we trace out spin 1 to establish a reduced density matrix for spins 2 and 3: $\rho_{f,23}$. Setting  $\rho_{i,23}=\ket{S}\bra{S}$, we compute the fidelity $f_{23}$ as above. Further details on the simulations are given in the Supplemental Material~\cite{supmat}.

\begin{table*}
	\centering\setcellgapes{4pt}\makegapedcells
	\begin{tabular}{|c|c|c|c|c|c|c|}
		\hline
		&\multicolumn{2}{c|}{$T_2^* = 18$ ns, $Q= 20$}&\multicolumn{2}{c|}{$T_2^* = 1000$ ns, $Q=20$} &\multicolumn{2}{c|}{$T_2^* = 1000$ ns, $Q=100$} \\ \hline
		& $1-f_1$& $1-f_{23}$ & $1-f_1$& $1-f_{23}$ & $1-f_1$& $1-f_{23}$  \\ \hline
		AQT & $1 \times 10^{-2}$ & $3 \times 10^{-2}$& $1 \times 10^{-3}$& $2 \times 10^{-3}$ & $6 \times 10^{-5}$ & $1 \times 10^{-4}$ \\ \hline
		$S_{12}S_{23}$     & $3 \times 10^{-2}$ & $1 \times 10^{-1}$ & $7 \times 10^{-3}$& $1 \times 10^{-2}$ & $5 \times 10^{-3}$ & $7 \times 10^{-3}$ \\ \hline

	\end{tabular}
	\caption{Simulated probability of incorrectly transferring either a single-spin eigenstate ($1-f_1$) or a singlet state ($1-f_{23}$) in a chain of three spins starting from the initial state $\ket{S_{12} \U_3}$ using AQT or SWAP gates. The sequence of SWAP gates that replicates the AQT is $S_{12}S_{23}$, where $S_{ij}$ indicates a SWAP between spins $i$ and $j$. The single-spin $T_2^*$ and exchange quality factors $Q$ are listed for three different cases. The case of $T_2^* = 18$ ns and $Q=20$ approximately corresponds to the experimental parameters studied here. $T_2^* = 1000$ ns and $Q=20$ correspond to what could likely be obtained in isotopically purified Si spin qubits. $T_2^* = 1000$ ns and $Q=100$ correspond to a significant reduction in both magnetic-field and charge noise over the experimental parameters studied here. In all cases, the errors associated with the AQT process are lower than the errors associated with the SWAP process. All values are rounded to one significant figure.}
	\label{tab1}
\end{table*}

Table~\ref{tab1} lists some of the results of these calculations. We find that both simulated probabilities exceed 0.95 for the experimental parameters studied here (single-spin $T_2^* \approx 18 $ ns and exchange quality factor $Q \approx 20$). Note that the experimentally observed probabilities are lower than this, as a result of SPAM errors. For example, the experimentally observed probability to measure a singlet in the right pair after two AQT steps, each with $T=100$ ns, is about 0.9, as shown in Fig.~\ref{fig4}. Based on additional measurements and simulations, we estimate that the state preparation fidelity associated with separated singlets or product states exceeds 0.95. We estimate that the charge-transfer sequence we use to implement a Pauli spin-blockade measurement has a fidelity of 0.94. The readout itself is characterized by fidelities associated with detector noise of 0.99 for the left pair and 0.95 for the right pair. Relaxation during readout contributes additional errors for excited states of 0.08 and 0.12 for the left and right pairs, respectively. Further details on SPAM errors are given in the Supplemental Material~\cite{supmat}.

We may extend this calculation to compute the expected transfer probability for single-qubit superposition states. We compute the transfer probability for $\ket{\phi}=\frac{1}{\sqrt{2}} (\ket{\U} +\ket{\D})$, as shown in Table~\ref{tab2}. We calculate that the probability to transfer this state from spin 3 to spin 1 can exceed 0.7, provided that spins 1 and 2 are initialized as $\ket{S}$. We calculate a similar probability for the other equal-superposition states. When spins 1 and 2 are initialized in states other than $\ket{S}$, the transfer probability is lower~\cite{supmat}.

\begin{table*}
	\centering\setcellgapes{4pt}\makegapedcells
	\begin{tabular}{|c|c|c|c|c|c|c|}
		\hline
		&\multicolumn{2}{c|}{$T_2^* = 18$ ns, $Q= 20$}&\multicolumn{2}{c|}{$T_2^* = 1000$ ns, $Q=20$} &\multicolumn{2}{c|}{$T_2^* = 1000$ ns, $Q=100$} \\ \hline
		& $1-f_1$& $1-f_{23}$ & $1-f_1$& $1-f_{23}$ & $1-f_1$& $1-f_{23}$  \\ \hline
		AQT & $2 \times 10^{-1}$ & $3 \times 10^{-2}$& $2 \times 10^{-3}$& $2 \times 10^{-3}$ & $8 \times 10^{-4}$ & $1 \times 10^{-4}$ \\ \hline
		$S_{12}S_{23}$     & $5\times 10^{-2}$ & $1 \times 10^{-1}$ & $7 \times 10^{-3}$& $1 \times 10^{-2}$ & $5 \times 10^{-3}$ & $7 \times 10^{-3}$ \\ \hline

	\end{tabular}
	\caption{Simulated probability of incorrectly transferring  a single-spin superposition state ($1-f_1$) and a singlet state ($1-f_{23}$) in a chain of three spins starting from the initial state $\frac{1}{\sqrt{2}}\ket{S_{12} (\U +\D)_3}$ using AQT or SWAP gates. The single-spin $T_2^*$ and exchange quality factors $Q$ are listed for the same three cases as above. In general, the AQT errors are lower than errors associated with SWAP gates, except when $T_2^*=18$ ns. In this case, the duration of the AQT significantly exceeds $T_2^*$. All values are rounded to one significant figure. }
	\label{tab2}
\end{table*}

We can also assess the expected transfer probability for the different states of spins 1 and 2, including the eigestates of the exchange operator. In this case, we define an initial state of the three-electron system as $\ket{\pi_{12}\phi_3}$, where $\ket{\pi}$ is a two-spin state. Supplementary Fig. 8 displays calculations of the transfer probabilities for different states $\ket{\pi}$. In general, the singlet state undergoes the highest-probability transfer. The origin of this advantage is evident in Fig.~\ref{fig2}. States containing the singlet are not degenerate with other states at any point during the evolution.

Finally, we may also simulate the process fidelity for transferring single-spin states via AQT. For electron spins in GaAs, the maximum simulated process fidelity to correctly transfer a single-spin state from dot 3 to dot 1 is about 0.7 at a total time of about 15 ns. This simulation agrees with the transfer probabilities listed above. It is not meaningful to ascribe a process fidelity for the transfer of the two-spin state, because AQT requires that the remaining two spins are in an eigenstate of exchange coupling at the beginning and end of the AQT. Spins 1 and 2 in the experiment of Fig.~\ref{fig3}(a) cannot, for example, have an arbitrary two-qubit state, similar to the requirement for an entangled state in conventional teleportation.

The primary limiting factor in these probabilities for GaAs quantum dots is the nuclear hyperfine gradient. First, the magnetic gradient limits the fidelity of the singlet state preparation. Second, a static magnetic gradient will tend to decrease the energy gaps in the system, requiring a slower pulse, or lowering the overall transfer fidelity for a pulse of the same speed~\cite{supmat}. Third, hyperfine fluctuations that quasistatically increase the magnetic gradients will also tend to decrease the adiabaticity during the pulse and lower the fidelity. 

In Si spin qubits, where nuclear hyperfine fields are suppressed, we expect that AQT can enable high-fidelity transfer of arbitrary states~\cite{Oh2013,gullans2020SCTAP,supmat}. For example, when $T_2^* > 1$ $\mu$s for single-spins, as is the case in isotopically purified Si, we simulate that transfer probabilities can exceed 0.99 for arbitrary single-qubit states (Tables~\ref{tab1} and \ref{tab2})~\cite{supmat}. We  simulate that process fidelities for single-spin transfer in silicon quantum dots can exceed 0.995. 

The AQT fidelity in our experiment is also affected by exchange-coupling calibration errors and charge noise, and this limitation will become more important in Si quantum dots.  Our current method of exchange-coupling control lets us set the couplings with an accuracy of about $10$ MHz~\cite{qiao2020}. Although we intend to ramp the exchange couplings linearly, errors in our exchange-coupling calibration can cause slight deviations from a linear ramp. In the future, more accurate modeling and control of exchange couplings should enable higher-fidelity state transfer. High-frequency charge noise can also have a similar effect. These deviations can reduce the overall fidelity, especially if the couplings are ramped more quickly than intended. We predict that high-frequency charge noise will be the limiting factor for AQT fidelities in Si quantum dots. The levels of quasi-static low-frequency charge noise in quantum dots have minimal effect on AQT, which is robust against small changes to the beginning and ending exchange-coupling values.  

\section{Discussion}

Our experiments show that AQT is a promising tool for quantum state transfer in semiconductor quantum-dot spin chains. Unlike methods for state transfer based on shuttling, AQT involves transferring quantum states without moving the qubits themselves, simplifying the process. 

Exchange-based SWAP gates can also transfer spin states without moving electrons~\cite{Kandel2019,Sigillito2019}. The simulation results shown in Table~\ref{tab1} indicate that AQT is more effective than a sequence of SWAP gates at transferring both eigenstates and spin singlets for the range of experimental parameters studied here. In particular, the SWAP sequence is vulnerable to errors associated with evolution of the singlet state in a magnetic gradient, although this evolution can in principle be corrected for via additional gates~\cite{Kandel2019}. Table~\ref{tab1} also suggests that the fluctuating hyperfine field is the dominant source of error, for both AQT and SWAP sequences, compared with exchange-coupling noise resulting from charge noise. 

In the case of superposition states, both AQT and the SWAP sequence are not very effective when $T_2^* \approx 18$ ns, as is the case in the experiments discussed here. In this case, AQT performs more poorly than the SWAP sequence, because the total duration of the AQT exceeds the $T_2^*$ time, in contrast to the SWAP sequence, which takes about 10 ns. A related challenge for the transmission of superposition states via AQT is the phase accrued during the relatively long gate times, if the quantum dots have different Zeeman splittings. This scenario can occur, for example, when micromagnets are used. However, this challenge can be solved with careful pulse calibration~\cite{Sigillito2019}. (The simulations presented in Tables~\ref{tab1} and \ref{tab2} were conducted with zero mean magnetic field in each dot to remove this constant phase.) We expect that in isotopically purified Si, superposition states can be transferred with high probability via AQT~\cite{gullans2020SCTAP}. In fact, our simulations indicate that AQT is more effective than SWAP gates at transferring superposition states in this case. We hypothesize that both high-frequency charge noise and residual errors due to hyperfine gradients are limiting factors for the short SWAP gates.  Encouragingly, the essential elements of the AQT process, including barrier-controlled exchange coupling~\cite{Reed2016SymSi,Takeda2020STsi} and Pauli spin-blockade readout~\cite{Jones2019,Elliot2020} are now common in Si spin qubits. 

In the future, state and process tomography, both requiring single-spin control and readout, will be required to definitively assess the performance of AQT for arbitrary qubit states. In particular, demonstrating an average single-qubit state transfer probability above the classical bound of 2/3~\cite{Massar1995}, or a process fidelity above 1/2~\cite{Steffen2013}, would verify the quantum-mechanical nature of this process. Measurements in addition to those presented here, which involve single-spin eigenstates, are needed to verify the quantum nature of the AQT process.

The AQT method implemented here is a highly robust method for the transfer of spin eigenstates and singlets in GaAs semiconductor quantum-dot arrays. The transfer of spin eigenstates is essential for readout in spin chains, and given the critical importance of spin singlets for various quantum information processing tasks, such as teleportation~\cite{qiao2020conditional} and superexchange~\cite{Oh2010,Oh2011}, it is likely that state-transfer via both AQT and SWAP gates will find use in spin-based quantum computing algorithms. We also expect that AQT will enable the high-fidelity transfer of arbitrary single-qubit states in Si spin qubits. However, the AQT process takes about ten times longer than a sequence of SWAP gates. Furthermore, while AQT transfers single-spin states between next-nearest-neighbor dots, SWAP gates transfer spin states between nearest-neighbor dots. From this point of view, the strengths of AQT complement the strengths of state transfer via SWAP gates. 

To conclude, our measurements provide evidence for adiabatic quantum-state transfer of both single-spin eigenstates and two-spin singlet states. We have also showed that the AQT protocol can be cascaded for efficient and robust quantum information transfer in a chain of semiconductor quantum-dot spin qubits. We believe that AQT will enable quantum state transfer in long chains of spin qubits for initialization, operation, and measurement in gate-based quantum computing architectures. An exciting prospect for future work is to harness many-body quantum states for direct, long-distance AQT~\cite{Chancellor2012,Farooq2015}. This work also opens up the possibility of adiabatic single qubit state- and gate-teleportation, as well as universal adiabatic quantum computing, in semiconductor quantum-dot spin qubits.

\section{Acknowledgments}
This work was sponsored by the Defense Advanced Research Projects Agency under Grant No. D18AC00025, the Army Research Office under Grant Nos. W911NF-16-1-0260 and W911NF-19-1-0167, and the National Science Foundation under Grant Nos. DMR-1941673 and DMR-2003287. The views and conclusions contained in this document are those of the authors and should not be interpreted as representing the official policies, either expressed or implied, of the Army Research Office or the U.S. Government. The U.S. Government is authorized to reproduce and distribute reprints for Government purposes notwithstanding any copyright notation herein.

\section{Author Contributions}
Y.P.K., H.Q., and J.M.N conceptualized the experiment and analyzed the data. Y.P.K and H.Q, conducted the investigation.  S.F., G.C.G., and M.J.M. provided resources. All authors participated in writing. J.M.N. supervised the effort.

\section{Competing Interests}
The authors declare no competing interests.

\section{Methods}
\subsection{Device}
Our quadruple quantum-dot device is fabricated on a GaAs/AlGaAs semiconductor heterostructure. The two-dimensional electron gas (2DEG) resides at the interface between the GaAs and AlGaAs layers, 91 nm below the surface of the wafer. The density and mobility of carriers in the 2DEG at a temperature of  $4$ K are $1.5\times 10^{11}$ cm$^{-2}$ and $2.5\times 10^{6}$ cm$^2$/Vs, respectively. Aluminum gates are arranged in a three-layer overlapping gate architecture and are fabricated using electron beam lithography. An additional top gate, not shown in Fig.~1(a) in the main text, covers all of the gates and the space around the center of device. Each of these metal gates are separated by a thin layer of native oxide formed on the gate surface. Voltages applied to the gates confine the electrons in the 2DEG. Each dot contains only one electron, and their chemical potentials are roughly the same, which we refer to as the symmetric configuration. The plunger and barrier gates are connected to arbitrary waveform generator channels via home made bias-tees. This configuration enables fast initialization, manipulation, and readout of the spins. Further details about the device are given in Ref.~\cite{Kandel2019}.

\subsection{Orthogonal control of the chemical potentials and exchange couplings}
Our device has four plunger gates $[p_1,p_2,p_3,p_4]$ for chemical potential control, three barrier gates $[b_1,b_2, b_3]$ for controlling the tunnel coupling between adjacent dots, and leads $[L_1, L_2]$ for controlling the system-environment interaction. In order to achieve individual control over the chemical potentials and exchange couplings, we define a set of virtual gates $\bm{G}=[P_1,P_2,P_3,P_4,B_1,B_2,B_3]^T$ as $\bm{G} = A\cdot\bm{g}$, where $\bm{g}=[p_1,p_2,p_3,p_4,b_1,b_2,b_3]^T$ is a set of physical gates, and $A$ is a $7\times7$ capacitance matrix~\cite{VanDiepen2018,Mills2019CA,qiao2020}. We achieve orthogonal control of the exchange couplings $\bm{J}=[J_{1},J_{2},J_{3}]$ by defining it as a non-linear function of the ``virtual" barrier gates using the Heitler-London model~\cite{qiao2020,Sousa2001}. Schematics of the virtual-gate pulses used to implement the AQT circuits shown in Figs. 3(a) and 4(a) in the main text are shown in Supplementary Figs.~1(a) and (b), respectively.

\subsection{State preparation and readout}
For initialization and readout, we configure the quadruple quantum-dot chain into two pairs. Dots 1 and 2 form the ``left" pair and dots 3 and 4 form the ``right" pair. We initialize the system in the (2, 0, 0, 2) charge state by lowering the chemical potentials of dots 1 and 4 below the Fermi level of the corresponding reservoir, while holding the chemical potentials of dots 2 and 3 above the Fermi level of that reservoir. The ground state of a pair of electrons in a single dot is the singlet state. We transfer one electron each from dots 1 and 4 into dots 2 and 3, respectively. Diabatic charge transfer maintains the joint spin states of the electrons, while adiabatic charge transfer prepares the electrons in spin eigenstates. We can also initialize either pair in the $\ket{\U\U}$ state via exchange of electrons from dots containing individual electron with the reservoir~\cite{Foletti2009,Orona2018Tp}. Measurement via Pauli spin blockade and safeguards to eliminate cross-talk are detailed in Refs.~\cite{Kandel2019,qiao2020}.

The state preparation estimates in the main text were obtained by initializing the dots in a particular state and measuring it. To estimate the singlet-state preparation fidelity on the left side, we load two electrons in dot 1 in the singlet state, transfer one of the electrons diabatically to dot 2, and project the spin-state of the electrons in dots 1 and 2 by diabatic transfer of the electron from dot 2 back to dot 1. Similarly, to estimate the fidelity of initializing the right pair in the ground state of the hyperfine field gradient, we load two electrons into dot 4 in the singlet state, transfer one of them adiabatically to dot 3, and then project the spin state of the electrons by adiabatic charge transfer of electron from dot 3 back to dot 4.

The experimental data of Figs. 3 and 4 in the main text involve measuring a singlet in the right pair. Generally, diabatic charge transfer together with a Pauli spin-blockade measurement suffice to measure a pair of electrons in the singlet-triplet basis~\cite{Petta2005,foletti2009universal}. However, the small inter-dot tunnel coupling limits the fidelity of diabatic projection in our device.  To measure a pair of electrons in the singlet-triplet basis in our device, we implemented a modified pulse sequence in which the electron pair is first evolved under the two-electron Hamiltonian
\begin{equation}
\label{JaRHamil}
H^{read}_{i,i+1} = J_{i}^{max} (1-t/\tau) \frac{h}{4} \bm{\sigma}_i\cdot\bm{\sigma}_{i+1} + \frac{h}{2}(B_i^z\sigma_i^z + B_{i+1}^z\sigma_{i+1}^z),
\end{equation}
where $\tau$ is the evolution time, and $J_i^{max}$ is the exchange coupling. To implement this Hamiltonian, we suddenly turn on a large exchange coupling between the two electrons, and slowly ramp it to zero. This procedure maps the singlet state to $\ket{\U \D}$ (or $\ket{\D \U}$, depending on the sign of the hyperfine gradient). Then we readout the electron pair by adiabatic projection, which remaps the state to the singlet-triplet basis. For $\tau=2\ \mu$s and $ J_{i}^{max}=300$ MHz, we estimate the fidelity of projection of spin state by this method to be 0.9 for the left side and 0.94 for the right side. A related method can be used to prepare entangled states in quantum-dot arrays~\cite{Farooq2015}. 

\subsection{Ground state of the magnetic field gradient}
To prepare the spin chain in a product state with $S^z=0$, we load two electrons in dots 1 and 4 each and transfer one electron from each of them to dots 2 and 3 adiabatically. The particular orientation of the spins in the chain after this step depends on the ground state of the hyperfine field gradient on both sides~\cite{Petta2005,foletti2009universal}. Since the hyperfine field fluctuates in time, the gradient also changes, and so does the ground-state spin configuration. Because our experiments involve preparing spin states on one side of the array and transferring them to the other side before the measurement, knowledge of the hyperfine configuration is critical. As we now discuss, we can monitor not only the sign of the gradients but also the ground-state spin configurations of the left and right sides in real time by measuring the evolution of the spin states in dots 2-3 under exchange coupling.

We define $f = \textnormal{sign}(B_2^z-B_1^z) \times \textnormal{sign}\left(B_4^z - B_3^z\right)$. In order to measure $f$, we initialize both sides as a product state with $S^z=0$, which we denote as the $GG$ configuration. We then evolve electrons 2 and 3 under exchange coupling for variable amount of time. When $f=+1$, the initial state of the chain is $\ket{\U_1\D_2\U_3\D_4}$ or $\ket{\D_1\U_2\D_3\U_4}$.  In these cases, the orientations of spins $2$ and $3$ are opposite, and they oscillate under exchange coupling. Adiabatic projection of the left and right sides, followed by measurement in the singlet/triplet basis, yields prominent exchange oscillations. But for $f=-1$, the initial state of the chain is $\ket{\U_1\D_2\D_3\U_4}$ or $\ket{\D_1\U_2\U_3\D_4}$. In these cases, the orientations of spins 2 and 3 are the same, and no exchange oscillations occur.

To determine the ground-state spin orientation, we load the left side in the $\ket{\U\U}$ state and the right side in a product state with $S^z=0$. We denote this as the $T^+G$ configuration, and we turn on exchange coupling between spins $2$ and $3$. For $\Delta B_{34}^z>0$, the spin state after initialization is $\ket{\U_1\U_2\D_3\U_4}$, and exchange oscillations between spins 2 and 3 can occur. However for $\Delta B_{34}^z<0$, the spin configuration of the chain after loading is $\ket{\U_1\U_2\U_3\D_4}$ and the spin states of electrons 2-3 do not evolve under exchange. The ground state spin-configuration in dots 1-2 can be inferred from the combined knowledge of $f$ and the spin configuration in dots 3-4. Supplementary Figure~2 illustrates these measurements.

\subsection{Post-selection of data}
The ground-state spin orientation of the spin chain was monitored during experiments by interleaving the pulses discussed above.  Specifically, we interleaved measurements of exchange between spins 2 and 3 with the $GG$ and $T^+G$ configurations (Supplementary Fig.~2). Here, $G$ denotes the two-spin ground state of the hyperfine field gradient in the $S^z=0$ subspace, and $T^+$ denotes spin pairs in the $\ket{\U\U}$ state. When spins 2-3 oscillate under exchange coupling with the $GG$ load, the ideal time-averaged singlet return probability on the right/left side is $\ta{P_S^{R/L}} = 0.5$ for $f=+1$. When $f=-1$ and the spins do not oscillate, $ \ta{P_S^{R/L}} = 1$. The measured values of $\ta{ P_S^{R/L}}$ may deviate from the ideal expectation due to a large $\Delta B_{23}^z$, load errors, or measurement errors. Thus, we define a threshold on the time-averaged singlet return probability as $P_{th}=0.75$, which is the mean of the two ideal values. We assume $f=+1$ when $\ta{P_S^{R/L}} <P_{th}$ and $f=-1$ when $\ta{P_S^{R/L}} >P_{th}$. 

The data associated with Figs.~3 and 4 in the main text consist of many repetitions. Each repetition, which consists of 512 or 256 single-shot measurements for Figs.~3 and 4, respectively, corresponding to each value of $T$, was assigned a value of $f$ based on the interleaved measurement discussed above. Individual repetitions with different values of $f$ are displayed in the main text. The data from all repetitions of the AQT experiments, as well as the average of the $f=\pm 1$ cases corresponding to Figs. 3, 4, and 5 of the main text are shown in Supplementary Figs. 3, 4, and 5, respectively. The averages across all $f=\pm1$ cases do not differ substantially from the individual repetitions displayed in the main text.
\subsection{Indirect AQT transfer probability assessment}
As an indirect qualitative indicator of the AQT transfer probability, we transferred states of the spin chain initialized in the state $\ket{S_{12}\U_3\D_4}$ or $\ket{S_{12}\D_3\U_4}$ via a cascade of AQT steps, with $T=127$~ns and varying $J^{max}$. Then, we evolved spins 1-2 and spins 3-4 under  exchange coupling for a variable amount of time [Supplementary Fig.~5(f)]. The measured singlet return probabilities on both sides were fitted to a function of the form $P_S^{R/L}(t)= V^{R/L} \cos(2\pi J_{i} t + \phi) \exp(-t^2/T_2^{*2}) + P_0 $, where $V^{R/L}$ is the visibility of exchange oscillations on the right/left side, $J_i$ is the frequency of exchange oscillations where $i=1$ for the left side and $i=3$ for the right side, $t$ is the evolution time, $\phi$ is a phase factor, $T_2^*$ is the dephasing time, and $P_0$ is the average of the time series. Here, $V^{R/L}$, $J_i$, $\phi$, $T_2^*$, and $P_0$ are fit parameters. In the adiabatic limit, spins 1 and 2 should be in a product state, and spins 3 and 4 should be in a singlet state at the end of the cascaded AQT sequence. As seen in Supplementary Fig.~5(g), $V^{L(R)}$ increases (decreases) in $J^{max}$ and finally saturates, as expected for adiabatic state transfer. For small values of $J^{max}$, $V^L$ is relatively small and $V^R$ is relatively large, due to the low AQT fidelity.  Apart from the infidelity of the AQT, the visibilities are limited by load and measurement errors, hyperfine noise, and charge noise.

\subsection{Simulation}
We accounted for known sources of errors and noise to reproduce experimental observations in the simulations. Load errors associated with the singlet, and the hyperfine ground state with $S^z=0$ were approximated by 
\begin{equation}\label{load}
\begin{aligned}
\ket{\tilde{S}} = s_1 \ket{S} + s_2\ket{T^0} +s_3 \ket{T^+} +s_4\ket{T^-}\\
\ket{\tilde{G}} = s_1 \ket{\U\D} + s_2\ket{\D\U} +s_3 \ket{T^+} +s_4\ket{T^-}
\end{aligned}
\end{equation}
where $\ket{S}=(\ket{\U\D}-\ket{\D\U})/\sqrt{2}$, $\ket{T^0}=(\ket{\U\D}+\ket{\D\U})/\sqrt{2}$, $\ket{T^+}=\ket{\U\U}$, and $\ket{T^-}=\ket{\D\D}$. $s_i$ is the probability amplitude of loading corresponding two-electron state. $f_S=|s_1|^2$, and $|s_2|^2=|s_3|^2=|s_4|^2=\frac{1-f_s}{3}$, where $f_S$ is singlet load fidelity. Based on measurements of the load process, we estimate that  $f_s=0.95$.

We define time dependent Hamiltonians acting on the spin chain 
\begin{equation}\label{H1}
\begin{aligned}
H_1(t)=J^{max} \frac{h}{4}\left[\left(1-\frac{t}{T}\right)\bm{\sigma}_1\cdot\bm{\sigma}_2+ \frac{t}{T} \bm{\sigma}_2\cdot\bm{\sigma}_3\right] + \frac{h}{2}\sum_{i=1}^{4} B_i^z\sigma_i^z,\\
H_2(t) = J^{max}\frac{h}{4}\left[\left(1-\frac{t}{T}\right) \bm{\sigma}_2\cdot\bm{\sigma}_3 + \frac{t}{T} \bm{\sigma}_3\cdot\bm{\sigma}_4\right]+ \frac{h}{2}\sum_{i=1}^{4} B_i^z\sigma_i^z.
\end{aligned}
\end{equation}
To simulate the AQT and SWAP experiment described in Fig. 3 in the main text, the spin chain was initialized in the state $\ket{\psi_0}=\ket{\tilde{S}}\otimes\ket{\tilde{G}}$. The state of the qubit chain after the AQT, $\ket{\psi_T}$, was obtained by numerical integration of the of the time-dependent Schrodinger equation:$$\ket{\psi_T} =U_B\prod_{j=0}^{N}{\exp(-iH_1(j\Delta t)\Delta t/\hbar)U_B\ket{\psi_0} },$$ where $N\times\Delta t = T$  and we used $\Delta t=1$ ns for all simulations. Here, $U_B=\exp(-i\pi \sum_{j=1}^{4} B_j^z\sigma_j^z t_{wait})$ is the evolution operator corresponding to the rise and fall time of the barrier pulses~\cite{qiao2020conditional}. We used $t_{wait}=1$ ns in all simulations. The SWAP operation between spin pair $j$ and $j+1$ was generated by a unitary operator $U_{j(j+1)}=\exp(-iH_{j (j+1)}^ST_S/\hbar)$, where $H_{j(j+1)}^S$ is
\begin{equation}\label{sHamiltonian}
H_{j(j+1)}^S= J_j\frac{h}{4}\bm{\sigma}_j\cdot\bm{\sigma}_{j+1} + \frac{h}{2}\sum_{j=1}^{4} B_j^z\sigma_j^z.
\end{equation}
The final state of the spin chain after the AQT and SWAP operations $U_{34}$ and $U_{23}$ is $\ket{\psi_f}=U_BU_{23}U_BU_{34}\ket{\psi_T}$. We used $J_i = 200$ MHz in experiments and simulation to generate the SWAP gates in the experiment of Fig. 3. The SWAP pulse time in the simulation is $T_S = \frac{1}{2 \sqrt{J_i^2 + \left(\Delta B^z_{i(i+1)}\right)^2}}$.


Similarly, for the simulation of the cascaded AQT experiment described in Fig. 4 of the main text, the initial state was set as $\ket{\psi_0}=\ket{\tilde{S}}\otimes\ket{\tilde{G}}$. The state of the spins after the AQT cascade was obtained by evaluating 
\begin{align}
\ket{\psi_f} &=U_B\prod_{k=0}^{N}{\exp(-iH_2(k\Delta t)\Delta t/\hbar)}\cdot \nonumber \\
&\left( \prod_{j=0}^{N}{\exp(-iH_1(j\Delta t)\Delta t/\hbar)U_B\ket{\psi_0}}\right) ,
\end{align}
where $N\times\Delta t = T$. The values of $T$ and $J^{max}$ were set to be the same as in the corresponding experiments. The target states for both experiments are $\ket{\varphi^R} = \ket{S}$, and 
\begin{equation}\label{phi}
\ket{\varphi^L} =
\begin{cases}
\ket{\U\D}, & \text{for} \Delta B_{34}^z<0\\
\ket{\D\U}, & \text{for} \Delta B_{34}^z>0 .
\end{cases}
\end{equation}
Finally, the singlet return probabilities were calculated as $P_S^{L/R} = \left| \braket{\varphi^{L/R}|\psi_f}\right|^2$.

The magnetic field in all simulations incorporates both the externally applied magnetic field of $0.5$ Tesla and the local hyperfine field. The values of the hyperfine field and its fluctuations were adjusted for better agreement between the simulation and the experimental data, and the specific values are presented in Supplementary Table~1. 

Charge noise directly affects the strength of exchange couplings. Low-frequency noise in the exchange couplings for each realization of the simulation was incorporated by sampling the exchange couplings from a Gaussian distribution with a target mean value $(J_0)$ and standard deviation equal to $J_0/(\sqrt{2}\pi Q)$. The exchange oscillation quality factors $Q$ for spin pairs 1-2, 2-3, and 3-4 are $\sim15$, $\sim15$, and $\sim20$, respectively. To incorporate high-frequency charge noise, we added an additional random error to the exchange couplings each ns, effectively adding white exchange-coupling noise with a bandwidth of about 1 GHz. The magnitude of this noise was chosen such that the exchange-oscillation quality factor is 10 times larger for a simulated Hahn-echo pulse sequence, as compared with a simulated free-induction decay pulse sequence. Although high-frequency charge noise has not been measured in GaAs quantum dots with overlapping gates, this assumption is consistent with the improvement in coherence times observed in Si devices with overlapping gates~\cite{Jock2018}. The simulated data were averaged over 256 different realizations. 

To include errors due to relaxation during the measurement, we define $g =1-\exp(-t_m/T_1)$ where $t_m$ is measurement time and $T_1$ is relaxation time. $g$ is the probability that the excited state  will relax to the ground state during readout. We define $1-f_r$ as the probability to misidentify the join spin state due to noise. The simulated return probability including relaxation and readout errors for either side is
\begin{equation}
\tilde{P}_S^{R/L}=  (2f_r-g-1) P_S^{R/L} + g -f_r+1.
\end{equation}
Specific values of $t_m$, $T_1$, and $f_r$ used in the simulations are 4 $\mu$s, 60 $\mu$s, and 0.99 for the left side, and 6 $\mu$s, 50 $\mu$s, and 0.95 for the right side, respectively. These values were obtained from the experiment.

\subsection{State-transfer fidelity estimation}
To estimate the probability of correctly transferring single-spin and two-spin states via a single AQT process, we simulate a three-spin system in the initial state $\ket{\pi_{12}\phi_3}$, where $\ket{\pi}$ is a two-qubit state, and $\ket{\phi}$ is a single-qubit state.  We numerically evolve this state in time under the Hamiltonian $H_1(t)$ defined in Eq.~(\ref{H1}) with $J^{max}=120$ MHz to a final state $\ket{\psi_f}$. We include noise and pulse errors in this simulation. To remove errors associated with phases between dots with different magnetic fields, we set the mean value of the magnetic field in each dot to zero, although we include hyperfine fluctuations as discussed above. We neglect SPAM errors in this calculation to assess the performance of the AQT operation itself. For each instance of the simulation, we trace out qubits 2 and 3 from the final state to obtain a reduced density matrix for qubit 1: $\rho_{f,1}$. Setting $\rho_{i,1}=\ket{\phi}\bra{\phi}$, we compute the state fidelity as $f_1=\left[\textnormal{Tr}\left(\sqrt{\sqrt{\rho_{f,1}}\rho_{i,1}\sqrt{\rho_{f,1}}}\right)\right]^2$. To estimate a transfer probability the initial two-spin state, we trace out qubit 1 to establish a reduced density matrix for qubits 2 and 3: $\rho_{f,23}$. Setting  $\rho_{i,23}=\ket{\pi}\bra{\pi}$, we compute the fidelity $f_{23}$ as above. 
We simulated interpolation times $0<T<256$ ns and chose the optimal value of $T$ for each different configuration. We averaged the simulation over 256 different low-frequency noise realizations. 

Supplementary Figures.~7,~8, and~9 show the simulated transfer probability for different initial states of spins 1-2 and spin 3 vs. single-spin $T_2^*$ values. Figure~7 suggests that arbitrary single-qubit states can be transferred with high fidelity in isotopically purified Si. Supplementary Figure.~8 also shows that while the singlet is the optimal configuration for spins 1-2, other eigenstates of exchange are expected to perform well in Si. When spins 1-2 are configured in superpositions of eigenstates of exchange (Supplementary Fig.~9), the AQT does not perform as well. Finally, the saturation of the error as  $T_2^*$ increases in Supplementary Figs.~7,~8, and~9 is related to the level of high-frequency charge noise we have included in the simulation. Thus, we expect that the ultimate limit on the AQT fidelity will be set by the levels of high-frequency charge noise in the system.

\subsection{Process fidelity simulation}
To estimate the process fidelity associated with transferring single-spin states via a single AQT process, we simulate a three-spin system in the initial state $\ket{S_{12}\phi_3}$, where $\ket{\phi}$ is any one of $\{\ket{\U},\ket{\D},\frac{1}{\sqrt{2}}\ket{\U+\D},\frac{1}{\sqrt{2}}\ket{\U+i\D}\}$  We numerically evolve this state in time under the Hamiltonian $H_1(t)$ defined in Eq.~(\ref{H1}) with $J^{max}=120$ MHz to a final state. We include noise and pulse errors in this simulation. To remove errors associated with phases between dots with different magnetic fields, we set the mean value of the magnetic field in each dot to zero, although we include hyperfine fluctuations as discussed above. We neglect SPAM errors in this calculation to assess the performance of the AQT operation itself. For each instance of the simulation, we trace out qubits 2 and 3 from the final state to obtain a reduced density matrix for qubit 1: $\rho_{f,1}$, and we define an effective input state $\rho_{i,1}=\ket{\phi}\bra{\phi}$, for each of the input states above. Using the four input states and simulated output states, we compute the process matrix $\chi$~\cite{chuang1997}. The ideal process matrix corresponds to the identity operation and has a single non-zero element, corresponding to evolution of the density matrix by the identity operator. For each instance of the simulation, we compute the process fidelity $Tr(\chi_{ideal}\chi)$, and then we average all fidelities for a given value of $T$ and noise parameters across all instances of the simulation. The simulations are averaged over 512 different realizations of low-frequency noise. We plot the resulting infidelities in Supplementary Fig.~10. 

\section{Data Availability}
The datasets generated during and/or analysed during the current study are available from the corresponding author on reasonable request.

%


\begin{thebibliography}{59}%
	\makeatletter
	\providecommand \@ifxundefined [1]{%
		\@ifx{#1\undefined}
	}%
	\providecommand \@ifnum [1]{%
		\ifnum #1\expandafter \@firstoftwo
		\else \expandafter \@secondoftwo
		\fi
	}%
	\providecommand \@ifx [1]{%
		\ifx #1\expandafter \@firstoftwo
		\else \expandafter \@secondoftwo
		\fi
	}%
	\providecommand \natexlab [1]{#1}%
	\providecommand \enquote  [1]{``#1''}%
	\providecommand \bibnamefont  [1]{#1}%
	\providecommand \bibfnamefont [1]{#1}%
	\providecommand \citenamefont [1]{#1}%
	\providecommand \href@noop [0]{\@secondoftwo}%
	\providecommand \href [0]{\begingroup \@sanitize@url \@href}%
	\providecommand \@href[1]{\@@startlink{#1}\@@href}%
	\providecommand \@@href[1]{\endgroup#1\@@endlink}%
	\providecommand \@sanitize@url [0]{\catcode `\\12\catcode `\$12\catcode
		`\&12\catcode `\#12\catcode `\^12\catcode `\_12\catcode `\%12\relax}%
	\providecommand \@@startlink[1]{}%
	\providecommand \@@endlink[0]{}%
	\providecommand \url  [0]{\begingroup\@sanitize@url \@url }%
	\providecommand \@url [1]{\endgroup\@href {#1}{\urlprefix }}%
	\providecommand \urlprefix  [0]{URL }%
	\providecommand \Eprint [0]{\href }%
	\providecommand \doibase [0]{http://dx.doi.org/}%
	\providecommand \selectlanguage [0]{\@gobble}%
	\providecommand \bibinfo  [0]{\@secondoftwo}%
	\providecommand \bibfield  [0]{\@secondoftwo}%
	\providecommand \translation [1]{[#1]}%
	\providecommand \BibitemOpen [0]{}%
	\providecommand \bibitemStop [0]{}%
	\providecommand \bibitemNoStop [0]{.\EOS\space}%
	\providecommand \EOS [0]{\spacefactor3000\relax}%
	\providecommand \BibitemShut  [1]{\csname bibitem#1\endcsname}%
	\let\auto@bib@innerbib\@empty
	\bibitem [{\citenamefont {Zajac}\ \emph {et~al.}(2016)\citenamefont {Zajac},
		\citenamefont {Hazard}, \citenamefont {Mi}, \citenamefont {Nielsen},\ and\
		\citenamefont {Petta}}]{Zajac2016}%
	\BibitemOpen
	\bibfield  {author} {\bibinfo {author} {\bibfnamefont {D.~M.}\ \bibnamefont
			{Zajac}}, \bibinfo {author} {\bibfnamefont {T.~M.}\ \bibnamefont {Hazard}},
		\bibinfo {author} {\bibfnamefont {X.}~\bibnamefont {Mi}}, \bibinfo {author}
		{\bibfnamefont {E.}~\bibnamefont {Nielsen}}, \ and\ \bibinfo {author}
		{\bibfnamefont {J.~R.}\ \bibnamefont {Petta}},\ }\bibfield  {title} {\enquote
		{\bibinfo {title} {Scalable gate architecture for a one-dimensional array of
				semiconductor spin qubits},}\ }\href@noop {} {\bibfield  {journal} {\bibinfo
			{journal} {Phys. Rev. Applied}\ }\textbf {\bibinfo {volume} {6}},\ \bibinfo
		{pages} {054013} (\bibinfo {year} {2016})}\BibitemShut {NoStop}%
	\bibitem [{\citenamefont {Volk}\ \emph {et~al.}(2019)\citenamefont {Volk},
		\citenamefont {Zwerver}, \citenamefont {Mukhopadhyay}, \citenamefont
		{Eendebak}, \citenamefont {van Diepen}, \citenamefont {Dehollain},
		\citenamefont {Hensgens}, \citenamefont {Fujita}, \citenamefont {Reichl},
		\citenamefont {Wegscheider},\ and\ \citenamefont {Vandersypen}}]{Volk2019}%
	\BibitemOpen
	\bibfield  {author} {\bibinfo {author} {\bibfnamefont {C.}~\bibnamefont
			{Volk}}, \bibinfo {author} {\bibfnamefont {A.~M.~J.}\ \bibnamefont
			{Zwerver}}, \bibinfo {author} {\bibfnamefont {U.}~\bibnamefont
			{Mukhopadhyay}}, \bibinfo {author} {\bibfnamefont {P.~T.}\ \bibnamefont
			{Eendebak}}, \bibinfo {author} {\bibfnamefont {C.~J.}\ \bibnamefont {van
				Diepen}}, \bibinfo {author} {\bibfnamefont {J.~P.}\ \bibnamefont
			{Dehollain}}, \bibinfo {author} {\bibfnamefont {T.}~\bibnamefont {Hensgens}},
		\bibinfo {author} {\bibfnamefont {T.}~\bibnamefont {Fujita}}, \bibinfo
		{author} {\bibfnamefont {C.}~\bibnamefont {Reichl}}, \bibinfo {author}
		{\bibfnamefont {W.}~\bibnamefont {Wegscheider}}, \ and\ \bibinfo {author}
		{\bibfnamefont {L.~M.~K.}\ \bibnamefont {Vandersypen}},\ }\bibfield  {title}
	{\enquote {\bibinfo {title} {Loading a quantum-dot based "qubyte"
				register},}\ }\href@noop {} {\bibfield  {journal} {\bibinfo  {journal} {npj
				Quantum Information}\ }\textbf {\bibinfo {volume} {5}},\ \bibinfo {pages}
		{29} (\bibinfo {year} {2019})}\BibitemShut {NoStop}%
	\bibitem [{\citenamefont {Baart}\ \emph {et~al.}(2016)\citenamefont {Baart},
		\citenamefont {Shafiei}, \citenamefont {Fujita}, \citenamefont {Reichl},
		\citenamefont {Wegscheider},\ and\ \citenamefont
		{Vandersypen}}]{Baart2016single}%
	\BibitemOpen
	\bibfield  {author} {\bibinfo {author} {\bibfnamefont {Timothy~Alexander}\
			\bibnamefont {Baart}}, \bibinfo {author} {\bibfnamefont {Mohammad}\
			\bibnamefont {Shafiei}}, \bibinfo {author} {\bibfnamefont {Takafumi}\
			\bibnamefont {Fujita}}, \bibinfo {author} {\bibfnamefont {Christian}\
			\bibnamefont {Reichl}}, \bibinfo {author} {\bibfnamefont {Werner}\
			\bibnamefont {Wegscheider}}, \ and\ \bibinfo {author} {\bibfnamefont {Lieven
				Mark~Koenraad}\ \bibnamefont {Vandersypen}},\ }\bibfield  {title} {\enquote
		{\bibinfo {title} {Single-spin ccd},}\ }\href@noop {} {\bibfield  {journal}
		{\bibinfo  {journal} {Nature nanotechnology}\ }\textbf {\bibinfo {volume}
			{11}},\ \bibinfo {pages} {330} (\bibinfo {year} {2016})}\BibitemShut
	{NoStop}%
	\bibitem [{\citenamefont {Mills}\ \emph
		{et~al.}(2019{\natexlab{a}})\citenamefont {Mills}, \citenamefont {Zajac},
		\citenamefont {Gullans}, \citenamefont {Schupp}, \citenamefont {Hazard},\
		and\ \citenamefont {Petta}}]{Mills2019CS}%
	\BibitemOpen
	\bibfield  {author} {\bibinfo {author} {\bibfnamefont {A.~R.}\ \bibnamefont
			{Mills}}, \bibinfo {author} {\bibfnamefont {D.~M.}\ \bibnamefont {Zajac}},
		\bibinfo {author} {\bibfnamefont {M.~J.}\ \bibnamefont {Gullans}}, \bibinfo
		{author} {\bibfnamefont {F.~J.}\ \bibnamefont {Schupp}}, \bibinfo {author}
		{\bibfnamefont {T.~M.}\ \bibnamefont {Hazard}}, \ and\ \bibinfo {author}
		{\bibfnamefont {J.~R.}\ \bibnamefont {Petta}},\ }\bibfield  {title} {\enquote
		{\bibinfo {title} {Shuttling a single charge across a one-dimensional array
				of silicon quantum dots},}\ }\href {\doibase 10.1038/s41467-019-08970-z}
	{\bibfield  {journal} {\bibinfo  {journal} {Nature Communications}\ }\textbf
		{\bibinfo {volume} {10}},\ \bibinfo {pages} {1063} (\bibinfo {year}
		{2019}{\natexlab{a}})}\BibitemShut {NoStop}%
	\bibitem [{\citenamefont {Hensgens}\ \emph {et~al.}(2017)\citenamefont
		{Hensgens}, \citenamefont {Fujita}, \citenamefont {Janssen}, \citenamefont
		{Li}, \citenamefont {Van~Diepen}, \citenamefont {Reichl}, \citenamefont
		{Wegscheider}, \citenamefont {Das~Sarma},\ and\ \citenamefont
		{Vandersypen}}]{Hensgens2017}%
	\BibitemOpen
	\bibfield  {author} {\bibinfo {author} {\bibfnamefont {T.}~\bibnamefont
			{Hensgens}}, \bibinfo {author} {\bibfnamefont {T.}~\bibnamefont {Fujita}},
		\bibinfo {author} {\bibfnamefont {L.}~\bibnamefont {Janssen}}, \bibinfo
		{author} {\bibfnamefont {Xiao}\ \bibnamefont {Li}}, \bibinfo {author}
		{\bibfnamefont {C.~J.}\ \bibnamefont {Van~Diepen}}, \bibinfo {author}
		{\bibfnamefont {C.}~\bibnamefont {Reichl}}, \bibinfo {author} {\bibfnamefont
			{W.}~\bibnamefont {Wegscheider}}, \bibinfo {author} {\bibfnamefont
			{S.}~\bibnamefont {Das~Sarma}}, \ and\ \bibinfo {author} {\bibfnamefont
			{L.~M.~K.}\ \bibnamefont {Vandersypen}},\ }\bibfield  {title} {\enquote
		{\bibinfo {title} {Quantum simulation of a fermi-hubbard model using a
				semiconductor quantum dot array},}\ }\href
	{https://doi.org/10.1038/nature23022} {\bibfield  {journal} {\bibinfo
			{journal} {Nature}\ }\textbf {\bibinfo {volume} {548}},\ \bibinfo {pages}
		{70--73} (\bibinfo {year} {2017})}\BibitemShut {NoStop}%
	\bibitem [{\citenamefont {van Diepen}\ \emph {et~al.}(2018)\citenamefont {van
			Diepen}, \citenamefont {Eendebak}, \citenamefont {Buijtendorp}, \citenamefont
		{Mukhopadhyay}, \citenamefont {Fujita}, \citenamefont {Reichl}, \citenamefont
		{Wegscheider},\ and\ \citenamefont {Vandersypen}}]{VanDiepen2018}%
	\BibitemOpen
	\bibfield  {author} {\bibinfo {author} {\bibfnamefont {C.~J.}\ \bibnamefont
			{van Diepen}}, \bibinfo {author} {\bibfnamefont {P.~T.}\ \bibnamefont
			{Eendebak}}, \bibinfo {author} {\bibfnamefont {B.~T.}\ \bibnamefont
			{Buijtendorp}}, \bibinfo {author} {\bibfnamefont {U.}~\bibnamefont
			{Mukhopadhyay}}, \bibinfo {author} {\bibfnamefont {T.}~\bibnamefont
			{Fujita}}, \bibinfo {author} {\bibfnamefont {C.}~\bibnamefont {Reichl}},
		\bibinfo {author} {\bibfnamefont {W.}~\bibnamefont {Wegscheider}}, \ and\
		\bibinfo {author} {\bibfnamefont {L.~M.~K.}\ \bibnamefont {Vandersypen}},\
	}\bibfield  {title} {\enquote {\bibinfo {title} {Automated tuning of
			inter-dot tunnel coupling in double quantum dots},}\ }\href {\doibase
	10.1063/1.5031034} {\bibfield  {journal} {\bibinfo  {journal} {Applied
			Physics Letters}\ }\textbf {\bibinfo {volume} {113}},\ \bibinfo {pages}
	{033101} (\bibinfo {year} {2018})}\BibitemShut {NoStop}%
\bibitem [{\citenamefont {Mills}\ \emph
	{et~al.}(2019{\natexlab{b}})\citenamefont {Mills}, \citenamefont {Feldman},
	\citenamefont {Monical}, \citenamefont {Lewis}, \citenamefont {Larson},
	\citenamefont {Mounce},\ and\ \citenamefont {Petta}}]{Mills2019CA}%
\BibitemOpen
\bibfield  {author} {\bibinfo {author} {\bibfnamefont {A.~R.}\ \bibnamefont
		{Mills}}, \bibinfo {author} {\bibfnamefont {M.~M.}\ \bibnamefont {Feldman}},
	\bibinfo {author} {\bibfnamefont {C.}~\bibnamefont {Monical}}, \bibinfo
	{author} {\bibfnamefont {P.~J.}\ \bibnamefont {Lewis}}, \bibinfo {author}
	{\bibfnamefont {K.~W.}\ \bibnamefont {Larson}}, \bibinfo {author}
	{\bibfnamefont {A.~M.}\ \bibnamefont {Mounce}}, \ and\ \bibinfo {author}
	{\bibfnamefont {J.~R.}\ \bibnamefont {Petta}},\ }\bibfield  {title} {\enquote
	{\bibinfo {title} {Computer-automated tuning procedures for semiconductor
			quantum dot arrays},}\ }\href {\doibase 10.1063/1.5121444} {\bibfield
	{journal} {\bibinfo  {journal} {Applied Physics Letters}\ }\textbf {\bibinfo
		{volume} {115}},\ \bibinfo {pages} {113501} (\bibinfo {year}
	{2019}{\natexlab{b}})}\BibitemShut {NoStop}%
\bibitem [{\citenamefont {Hsiao}\ \emph {et~al.}(2020)\citenamefont {Hsiao},
	\citenamefont {van Diepen}, \citenamefont {Mukhopadhyay}, \citenamefont
	{Reichl}, \citenamefont {Wegscheider},\ and\ \citenamefont
	{Vandersypen}}]{Hsiao2020}%
\BibitemOpen
\bibfield  {author} {\bibinfo {author} {\bibfnamefont {T.~K.}\ \bibnamefont
		{Hsiao}}, \bibinfo {author} {\bibfnamefont {C.~J.}\ \bibnamefont {van
			Diepen}}, \bibinfo {author} {\bibfnamefont {U.}~\bibnamefont {Mukhopadhyay}},
	\bibinfo {author} {\bibfnamefont {C.}~\bibnamefont {Reichl}}, \bibinfo
	{author} {\bibfnamefont {W.}~\bibnamefont {Wegscheider}}, \ and\ \bibinfo
	{author} {\bibfnamefont {L.~M.~K.}\ \bibnamefont {Vandersypen}},\ }\bibfield
{title} {\enquote {\bibinfo {title} {Efficient orthogonal control of tunnel
			couplings in a quantum dot array},}\ }\href@noop {} {\  (\bibinfo {year}
	{2020})},\ \Eprint {http://arxiv.org/abs/2001.07671} {arXiv:2001.07671}
\BibitemShut {NoStop}%
\bibitem [{\citenamefont {Zwolak}\ \emph {et~al.}(2020)\citenamefont {Zwolak},
	\citenamefont {McJunkin}, \citenamefont {Kalantre}, \citenamefont {Dodson},
	\citenamefont {MacQuarrie}, \citenamefont {Savage}, \citenamefont {Lagally},
	\citenamefont {Coppersmith}, \citenamefont {Eriksson},\ and\ \citenamefont
	{Taylor}}]{zwolak2020autotuning}%
\BibitemOpen
\bibfield  {author} {\bibinfo {author} {\bibfnamefont {Justyna~P}\
		\bibnamefont {Zwolak}}, \bibinfo {author} {\bibfnamefont {Thomas}\
		\bibnamefont {McJunkin}}, \bibinfo {author} {\bibfnamefont {Sandesh~S}\
		\bibnamefont {Kalantre}}, \bibinfo {author} {\bibfnamefont {JP}~\bibnamefont
		{Dodson}}, \bibinfo {author} {\bibfnamefont {ER}~\bibnamefont {MacQuarrie}},
	\bibinfo {author} {\bibfnamefont {DE}~\bibnamefont {Savage}}, \bibinfo
	{author} {\bibfnamefont {MG}~\bibnamefont {Lagally}}, \bibinfo {author}
	{\bibfnamefont {SN}~\bibnamefont {Coppersmith}}, \bibinfo {author}
	{\bibfnamefont {Mark~A}\ \bibnamefont {Eriksson}}, \ and\ \bibinfo {author}
	{\bibfnamefont {Jacob~M}\ \bibnamefont {Taylor}},\ }\bibfield  {title}
{\enquote {\bibinfo {title} {Autotuning of double-dot devices in situ with
			machine learning},}\ }\href@noop {} {\bibfield  {journal} {\bibinfo
		{journal} {Physical Review Applied}\ }\textbf {\bibinfo {volume} {13}},\
	\bibinfo {pages} {034075} (\bibinfo {year} {2020})}\BibitemShut {NoStop}%
\bibitem [{\citenamefont {Qiao}\ \emph
	{et~al.}(2020{\natexlab{a}})\citenamefont {Qiao}, \citenamefont {Kandel},
	\citenamefont {Deng}, \citenamefont {Fallahi}, \citenamefont {Gardner},
	\citenamefont {Manfra}, \citenamefont {Barnes},\ and\ \citenamefont
	{Nichol}}]{qiao2020}%
\BibitemOpen
\bibfield  {author} {\bibinfo {author} {\bibfnamefont {Haifeng}\ \bibnamefont
		{Qiao}}, \bibinfo {author} {\bibfnamefont {Yadav~P.}\ \bibnamefont {Kandel}},
	\bibinfo {author} {\bibfnamefont {Kuangyin}\ \bibnamefont {Deng}}, \bibinfo
	{author} {\bibfnamefont {Saeed}\ \bibnamefont {Fallahi}}, \bibinfo {author}
	{\bibfnamefont {Geoffrey~C.}\ \bibnamefont {Gardner}}, \bibinfo {author}
	{\bibfnamefont {Michael~J.}\ \bibnamefont {Manfra}}, \bibinfo {author}
	{\bibfnamefont {Edwin}\ \bibnamefont {Barnes}}, \ and\ \bibinfo {author}
	{\bibfnamefont {John~M.}\ \bibnamefont {Nichol}},\ }\bibfield  {title}
{\enquote {\bibinfo {title} {Coherent multi-spin exchange in a quantum-dot
			spin chain},}\ }\href@noop {} {\  (\bibinfo {year} {2020}{\natexlab{a}})},\
\Eprint {http://arxiv.org/abs/2001.02277} {arXiv:2001.02277} \BibitemShut
{NoStop}%
\bibitem [{\citenamefont {Kandel}\ \emph {et~al.}(2019)\citenamefont {Kandel},
	\citenamefont {Qiao}, \citenamefont {Fallahi}, \citenamefont {Gardner},
	\citenamefont {Manfra},\ and\ \citenamefont {Nichol}}]{Kandel2019}%
\BibitemOpen
\bibfield  {author} {\bibinfo {author} {\bibfnamefont {Yadav~P.}\
		\bibnamefont {Kandel}}, \bibinfo {author} {\bibfnamefont {Haifeng}\
		\bibnamefont {Qiao}}, \bibinfo {author} {\bibfnamefont {Saeed}\ \bibnamefont
		{Fallahi}}, \bibinfo {author} {\bibfnamefont {Geoffrey~C.}\ \bibnamefont
		{Gardner}}, \bibinfo {author} {\bibfnamefont {Michael~J.}\ \bibnamefont
		{Manfra}}, \ and\ \bibinfo {author} {\bibfnamefont {John~M.}\ \bibnamefont
		{Nichol}},\ }\bibfield  {title} {\enquote {\bibinfo {title} {Coherent
			spin-state transfer via heisenberg exchange},}\ }\href@noop {} {\bibfield
	{journal} {\bibinfo  {journal} {Nature}\ }\textbf {\bibinfo {volume} {573}},\
	\bibinfo {pages} {553--557} (\bibinfo {year} {2019})}\BibitemShut {NoStop}%
\bibitem [{\citenamefont {Qiao}\ \emph
	{et~al.}(2020{\natexlab{b}})\citenamefont {Qiao}, \citenamefont {Kandel},
	\citenamefont {Manikandan}, \citenamefont {Jordan}, \citenamefont {Fallahi},
	\citenamefont {Gardner}, \citenamefont {Manfra},\ and\ \citenamefont
	{Nichol}}]{qiao2020conditional}%
\BibitemOpen
\bibfield  {author} {\bibinfo {author} {\bibfnamefont {Haifeng}\ \bibnamefont
		{Qiao}}, \bibinfo {author} {\bibfnamefont {Yadav~P}\ \bibnamefont {Kandel}},
	\bibinfo {author} {\bibfnamefont {Sreenath~K}\ \bibnamefont {Manikandan}},
	\bibinfo {author} {\bibfnamefont {Andrew~N}\ \bibnamefont {Jordan}}, \bibinfo
	{author} {\bibfnamefont {Saeed}\ \bibnamefont {Fallahi}}, \bibinfo {author}
	{\bibfnamefont {Geoffrey~C}\ \bibnamefont {Gardner}}, \bibinfo {author}
	{\bibfnamefont {Michael~J}\ \bibnamefont {Manfra}}, \ and\ \bibinfo {author}
	{\bibfnamefont {John~M}\ \bibnamefont {Nichol}},\ }\bibfield  {title}
{\enquote {\bibinfo {title} {Conditional teleportation of quantum-dot spin
			states},}\ }\href@noop {} {\bibfield  {journal} {\bibinfo  {journal} {Nature
			Communications}\ }\textbf {\bibinfo {volume} {11}},\ \bibinfo {pages} {1--9}
	(\bibinfo {year} {2020}{\natexlab{b}})}\BibitemShut {NoStop}%
\bibitem [{\citenamefont {Fujita}\ \emph {et~al.}(2017)\citenamefont {Fujita},
	\citenamefont {Baart}, \citenamefont {Reichl}, \citenamefont {Wegscheider},\
	and\ \citenamefont {Vandersypen}}]{Fujita2017}%
\BibitemOpen
\bibfield  {author} {\bibinfo {author} {\bibfnamefont {Takafumi}\
		\bibnamefont {Fujita}}, \bibinfo {author} {\bibfnamefont {Timothy~Alexander}\
		\bibnamefont {Baart}}, \bibinfo {author} {\bibfnamefont {Christian}\
		\bibnamefont {Reichl}}, \bibinfo {author} {\bibfnamefont {Werner}\
		\bibnamefont {Wegscheider}}, \ and\ \bibinfo {author} {\bibfnamefont {Lieven
			Mark~Koenraad}\ \bibnamefont {Vandersypen}},\ }\bibfield  {title} {\enquote
	{\bibinfo {title} {Coherent shuttle of electron-spin states},}\ }\href
{\doibase 10.1038/s41534-017-0024-4} {\bibfield  {journal} {\bibinfo
		{journal} {npj Quantum Information}\ }\textbf {\bibinfo {volume} {3}},\
	\bibinfo {pages} {22} (\bibinfo {year} {2017})}\BibitemShut {NoStop}%
\bibitem [{\citenamefont {Flentje}\ \emph {et~al.}(2017)\citenamefont
	{Flentje}, \citenamefont {Mortemousque}, \citenamefont {Thalineau},
	\citenamefont {Ludwig}, \citenamefont {Wieck}, \citenamefont {B{\"a}uerle},\
	and\ \citenamefont {Meunier}}]{Flentje2017coherent}%
\BibitemOpen
\bibfield  {author} {\bibinfo {author} {\bibfnamefont {H}~\bibnamefont
		{Flentje}}, \bibinfo {author} {\bibfnamefont {P-A}\ \bibnamefont
		{Mortemousque}}, \bibinfo {author} {\bibfnamefont {R}~\bibnamefont
		{Thalineau}}, \bibinfo {author} {\bibfnamefont {A}~\bibnamefont {Ludwig}},
	\bibinfo {author} {\bibfnamefont {AD}~\bibnamefont {Wieck}}, \bibinfo
	{author} {\bibfnamefont {C}~\bibnamefont {B{\"a}uerle}}, \ and\ \bibinfo
	{author} {\bibfnamefont {T}~\bibnamefont {Meunier}},\ }\bibfield  {title}
{\enquote {\bibinfo {title} {Coherent long-distance displacement of
			individual electron spins},}\ }\href@noop {} {\bibfield  {journal} {\bibinfo
		{journal} {Nature communications}\ }\textbf {\bibinfo {volume} {8}},\
	\bibinfo {pages} {1--6} (\bibinfo {year} {2017})}\BibitemShut {NoStop}%
\bibitem [{\citenamefont {Nakajima}\ \emph {et~al.}(2018)\citenamefont
	{Nakajima}, \citenamefont {Delbecq}, \citenamefont {Otsuka}, \citenamefont
	{Amaha}, \citenamefont {Yoneda}, \citenamefont {Noiri}, \citenamefont
	{Takeda}, \citenamefont {Allison}, \citenamefont {Ludwig}, \citenamefont
	{Wieck}, \citenamefont {Hu}, \citenamefont {Nori},\ and\ \citenamefont
	{Tarucha}}]{Nakajima2018}%
\BibitemOpen
\bibfield  {author} {\bibinfo {author} {\bibfnamefont {Takashi}\ \bibnamefont
		{Nakajima}}, \bibinfo {author} {\bibfnamefont {Matthieu~R.}\ \bibnamefont
		{Delbecq}}, \bibinfo {author} {\bibfnamefont {Tomohiro}\ \bibnamefont
		{Otsuka}}, \bibinfo {author} {\bibfnamefont {Shinichi}\ \bibnamefont
		{Amaha}}, \bibinfo {author} {\bibfnamefont {Jun}\ \bibnamefont {Yoneda}},
	\bibinfo {author} {\bibfnamefont {Akito}\ \bibnamefont {Noiri}}, \bibinfo
	{author} {\bibfnamefont {Kenta}\ \bibnamefont {Takeda}}, \bibinfo {author}
	{\bibfnamefont {Giles}\ \bibnamefont {Allison}}, \bibinfo {author}
	{\bibfnamefont {Arne}\ \bibnamefont {Ludwig}}, \bibinfo {author}
	{\bibfnamefont {Andreas~D.}\ \bibnamefont {Wieck}}, \bibinfo {author}
	{\bibfnamefont {Xuedong}\ \bibnamefont {Hu}}, \bibinfo {author}
	{\bibfnamefont {Franco}\ \bibnamefont {Nori}}, \ and\ \bibinfo {author}
	{\bibfnamefont {Seigo}\ \bibnamefont {Tarucha}},\ }\bibfield  {title}
{\enquote {\bibinfo {title} {Coherent transfer of electron spin correlations
			assisted by dephasing noise},}\ }\href {\doibase 10.1038/s41467-018-04544-7}
{\bibfield  {journal} {\bibinfo  {journal} {Nature Communications}\ }\textbf
	{\bibinfo {volume} {9}},\ \bibinfo {pages} {2133} (\bibinfo {year}
	{2018})}\BibitemShut {NoStop}%
\bibitem [{\citenamefont {Bertrand}\ \emph {et~al.}(2016)\citenamefont
	{Bertrand}, \citenamefont {Hermelin}, \citenamefont {Takada}, \citenamefont
	{Yamamoto}, \citenamefont {Tarucha}, \citenamefont {Ludwig}, \citenamefont
	{Wieck}, \citenamefont {B{\"a}uerle},\ and\ \citenamefont
	{Meunier}}]{bertrand2016fast}%
\BibitemOpen
\bibfield  {author} {\bibinfo {author} {\bibfnamefont {Benoit}\ \bibnamefont
		{Bertrand}}, \bibinfo {author} {\bibfnamefont {Sylvain}\ \bibnamefont
		{Hermelin}}, \bibinfo {author} {\bibfnamefont {Shintaro}\ \bibnamefont
		{Takada}}, \bibinfo {author} {\bibfnamefont {Michihisa}\ \bibnamefont
		{Yamamoto}}, \bibinfo {author} {\bibfnamefont {Seigo}\ \bibnamefont
		{Tarucha}}, \bibinfo {author} {\bibfnamefont {Arne}\ \bibnamefont {Ludwig}},
	\bibinfo {author} {\bibfnamefont {Andreas~D}\ \bibnamefont {Wieck}}, \bibinfo
	{author} {\bibfnamefont {Christopher}\ \bibnamefont {B{\"a}uerle}}, \ and\
	\bibinfo {author} {\bibfnamefont {Tristan}\ \bibnamefont {Meunier}},\
}\bibfield  {title} {\enquote {\bibinfo {title} {Fast spin information
		transfer between distant quantum dots using individual electrons},}\
}\href@noop {} {\bibfield  {journal} {\bibinfo  {journal} {Nature
		nanotechnology}\ }\textbf {\bibinfo {volume} {11}},\ \bibinfo {pages} {672}
(\bibinfo {year} {2016})}\BibitemShut {NoStop}%
\bibitem [{\citenamefont {Sigillito}\ \emph {et~al.}(2019)\citenamefont
	{Sigillito}, \citenamefont {Gullans}, \citenamefont {Edge}, \citenamefont
	{Borselli},\ and\ \citenamefont {Petta}}]{Sigillito2019}%
\BibitemOpen
\bibfield  {author} {\bibinfo {author} {\bibfnamefont {A.~J.}\ \bibnamefont
		{Sigillito}}, \bibinfo {author} {\bibfnamefont {M.~J.}\ \bibnamefont
		{Gullans}}, \bibinfo {author} {\bibfnamefont {L.~F.}\ \bibnamefont {Edge}},
	\bibinfo {author} {\bibfnamefont {M.}~\bibnamefont {Borselli}}, \ and\
	\bibinfo {author} {\bibfnamefont {J.~R.}\ \bibnamefont {Petta}},\ }\bibfield
{title} {\enquote {\bibinfo {title} {Coherent transfer of quantum information
			in a silicon double quantum dot using resonant swap gates},}\ }\href
{\doibase 10.1038/s41534-019-0225-0} {\bibfield  {journal} {\bibinfo
		{journal} {npj Quantum Information}\ }\textbf {\bibinfo {volume} {5}},\
	\bibinfo {pages} {110} (\bibinfo {year} {2019})}\BibitemShut {NoStop}%
\bibitem [{\citenamefont {Baart}\ \emph {et~al.}(2017)\citenamefont {Baart},
	\citenamefont {Fujita}, \citenamefont {Reichl}, \citenamefont {Wegscheider},\
	and\ \citenamefont {Vandersypen}}]{Baart2017}%
\BibitemOpen
\bibfield  {author} {\bibinfo {author} {\bibfnamefont {Timothy~Alexander}\
		\bibnamefont {Baart}}, \bibinfo {author} {\bibfnamefont {Takafumi}\
		\bibnamefont {Fujita}}, \bibinfo {author} {\bibfnamefont {Christian}\
		\bibnamefont {Reichl}}, \bibinfo {author} {\bibfnamefont {Werner}\
		\bibnamefont {Wegscheider}}, \ and\ \bibinfo {author} {\bibfnamefont {Lieven
			Mark~Koenraad}\ \bibnamefont {Vandersypen}},\ }\bibfield  {title} {\enquote
	{\bibinfo {title} {Coherent spin-exchange via a quantum mediator},}\
}\href@noop {} {\bibfield  {journal} {\bibinfo  {journal} {Nature
		Nanotechnology}\ }\textbf {\bibinfo {volume} {12}},\ \bibinfo {pages}
{26--30} (\bibinfo {year} {2017})}\BibitemShut {NoStop}%
\bibitem [{\citenamefont {Malinowski}\ \emph {et~al.}(2019)\citenamefont
	{Malinowski}, \citenamefont {Martins}, \citenamefont {Smith}, \citenamefont
	{Bartlett}, \citenamefont {Doherty}, \citenamefont {Nissen}, \citenamefont
	{Fallahi}, \citenamefont {Gardner}, \citenamefont {Manfra}, \citenamefont
	{Marcus},\ and\ \citenamefont {Kuemmeth}}]{Malinowski2019}%
\BibitemOpen
\bibfield  {author} {\bibinfo {author} {\bibfnamefont {Filip~K.}\
		\bibnamefont {Malinowski}}, \bibinfo {author} {\bibfnamefont {Frederico}\
		\bibnamefont {Martins}}, \bibinfo {author} {\bibfnamefont {Thomas~B.}\
		\bibnamefont {Smith}}, \bibinfo {author} {\bibfnamefont {Stephen~D.}\
		\bibnamefont {Bartlett}}, \bibinfo {author} {\bibfnamefont {Andrew~C.}\
		\bibnamefont {Doherty}}, \bibinfo {author} {\bibfnamefont {Peter~D.}\
		\bibnamefont {Nissen}}, \bibinfo {author} {\bibfnamefont {Saeed}\
		\bibnamefont {Fallahi}}, \bibinfo {author} {\bibfnamefont {Geoffrey~C.}\
		\bibnamefont {Gardner}}, \bibinfo {author} {\bibfnamefont {Michael~J.}\
		\bibnamefont {Manfra}}, \bibinfo {author} {\bibfnamefont {Charles~M.}\
		\bibnamefont {Marcus}}, \ and\ \bibinfo {author} {\bibfnamefont {Ferdinand}\
		\bibnamefont {Kuemmeth}},\ }\bibfield  {title} {\enquote {\bibinfo {title}
		{Fast spin exchange across a multielectron mediator},}\ }\href {\doibase
	10.1038/s41467-019-09194-x} {\bibfield  {journal} {\bibinfo  {journal}
		{Nature Communications}\ }\textbf {\bibinfo {volume} {10}},\ \bibinfo {pages}
	{1196} (\bibinfo {year} {2019})}\BibitemShut {NoStop}%
\bibitem [{\citenamefont {Farhi}\ \emph {et~al.}(2000)\citenamefont {Farhi},
	\citenamefont {Goldstone}, \citenamefont {Gutmann},\ and\ \citenamefont
	{Sipser}}]{farhi2000}%
\BibitemOpen
\bibfield  {author} {\bibinfo {author} {\bibfnamefont {Edward}\ \bibnamefont
		{Farhi}}, \bibinfo {author} {\bibfnamefont {Jeffrey}\ \bibnamefont
		{Goldstone}}, \bibinfo {author} {\bibfnamefont {Sam}\ \bibnamefont
		{Gutmann}}, \ and\ \bibinfo {author} {\bibfnamefont {Michael}\ \bibnamefont
		{Sipser}},\ }\bibfield  {title} {\enquote {\bibinfo {title} {Quantum
			computation by adiabatic evolution},}\ }\href@noop {} {\  (\bibinfo {year}
	{2000})},\ \Eprint {http://arxiv.org/abs/quant-ph/0001106}
{arXiv:quant-ph/0001106} \BibitemShut {NoStop}%
\bibitem [{\citenamefont {Bacon}\ and\ \citenamefont
	{Flammia}(2009)}]{Bacon2009AGT}%
\BibitemOpen
\bibfield  {author} {\bibinfo {author} {\bibfnamefont {Dave}\ \bibnamefont
		{Bacon}}\ and\ \bibinfo {author} {\bibfnamefont {Steven~T.}\ \bibnamefont
		{Flammia}},\ }\bibfield  {title} {\enquote {\bibinfo {title} {Adiabatic gate
			teleportation},}\ }\href {\doibase 10.1103/PhysRevLett.103.120504} {\bibfield
	{journal} {\bibinfo  {journal} {Phys. Rev. Lett.}\ }\textbf {\bibinfo
		{volume} {103}},\ \bibinfo {pages} {120504} (\bibinfo {year}
	{2009})}\BibitemShut {NoStop}%
\bibitem [{\citenamefont {Greentree}\ \emph {et~al.}(2004)\citenamefont
	{Greentree}, \citenamefont {Cole}, \citenamefont {Hamilton},\ and\
	\citenamefont {Hollenberg}}]{Lloyd2004}%
\BibitemOpen
\bibfield  {author} {\bibinfo {author} {\bibfnamefont {Andrew~D.}\
		\bibnamefont {Greentree}}, \bibinfo {author} {\bibfnamefont {Jared~H.}\
		\bibnamefont {Cole}}, \bibinfo {author} {\bibfnamefont {A.~R.}\ \bibnamefont
		{Hamilton}}, \ and\ \bibinfo {author} {\bibfnamefont {Lloyd C.~L.}\
		\bibnamefont {Hollenberg}},\ }\bibfield  {title} {\enquote {\bibinfo {title}
		{Coherent electronic transfer in quantum dot systems using adiabatic
			passage},}\ }\href@noop {} {\bibfield  {journal} {\bibinfo  {journal} {Phys.
			Rev. B}\ }\textbf {\bibinfo {volume} {70}},\ \bibinfo {pages} {235317}
	(\bibinfo {year} {2004})}\BibitemShut {NoStop}%
\bibitem [{\citenamefont {Srinivasa}\ \emph
	{et~al.}(2007{\natexlab{a}})\citenamefont {Srinivasa}, \citenamefont {Levy},\
	and\ \citenamefont {Hellberg}}]{Srinivasa2007}%
\BibitemOpen
\bibfield  {author} {\bibinfo {author} {\bibfnamefont {Vanita}\ \bibnamefont
		{Srinivasa}}, \bibinfo {author} {\bibfnamefont {Jeremy}\ \bibnamefont
		{Levy}}, \ and\ \bibinfo {author} {\bibfnamefont {C.~Stephen}\ \bibnamefont
		{Hellberg}},\ }\bibfield  {title} {\enquote {\bibinfo {title} {Flying spin
			qubits: A method for encoding and transporting qubits within a dimerized
			heisenberg spin-$\frac{1}{2}$ chain},}\ }\href {\doibase
	10.1103/PhysRevB.76.094411} {\bibfield  {journal} {\bibinfo  {journal} {Phys.
			Rev. B}\ }\textbf {\bibinfo {volume} {76}},\ \bibinfo {pages} {094411}
	(\bibinfo {year} {2007}{\natexlab{a}})}\BibitemShut {NoStop}%
\bibitem [{\citenamefont {Srinivasa}\ \emph
	{et~al.}(2007{\natexlab{b}})\citenamefont {Srinivasa}, \citenamefont {Levy},\
	and\ \citenamefont {Hellberg}}]{Srinivasa2009}%
\BibitemOpen
\bibfield  {author} {\bibinfo {author} {\bibfnamefont {Vanita}\ \bibnamefont
		{Srinivasa}}, \bibinfo {author} {\bibfnamefont {Jeremy}\ \bibnamefont
		{Levy}}, \ and\ \bibinfo {author} {\bibfnamefont {C.~Stephen}\ \bibnamefont
		{Hellberg}},\ }\bibfield  {title} {\enquote {\bibinfo {title} {Flying spin
			qubits: A method for encoding and transporting qubits within a dimerized
			heisenberg spin-$\frac{1}{2}$ chain},}\ }\href {\doibase
	10.1103/PhysRevB.76.094411} {\bibfield  {journal} {\bibinfo  {journal} {Phys.
			Rev. B}\ }\textbf {\bibinfo {volume} {76}},\ \bibinfo {pages} {094411}
	(\bibinfo {year} {2007}{\natexlab{b}})}\BibitemShut {NoStop}%
\bibitem [{\citenamefont {Oh}\ \emph {et~al.}(2013)\citenamefont {Oh},
	\citenamefont {Shim}, \citenamefont {Fei}, \citenamefont {Friesen},\ and\
	\citenamefont {Hu}}]{Oh2013}%
\BibitemOpen
\bibfield  {author} {\bibinfo {author} {\bibfnamefont {Sangchul}\
		\bibnamefont {Oh}}, \bibinfo {author} {\bibfnamefont {Yun-Pil}\ \bibnamefont
		{Shim}}, \bibinfo {author} {\bibfnamefont {Jianjia}\ \bibnamefont {Fei}},
	\bibinfo {author} {\bibfnamefont {Mark}\ \bibnamefont {Friesen}}, \ and\
	\bibinfo {author} {\bibfnamefont {Xuedong}\ \bibnamefont {Hu}},\ }\bibfield
{title} {\enquote {\bibinfo {title} {Resonant adiabatic passage with three
			qubits},}\ }\href {\doibase 10.1103/PhysRevA.87.022332} {\bibfield  {journal}
	{\bibinfo  {journal} {Phys. Rev. A}\ }\textbf {\bibinfo {volume} {87}},\
	\bibinfo {pages} {022332} (\bibinfo {year} {2013})}\BibitemShut {NoStop}%
\bibitem [{\citenamefont {Menchon-Enrich}\ \emph {et~al.}(2016)\citenamefont
	{Menchon-Enrich}, \citenamefont {Benseny}, \citenamefont {Ahufinger},
	\citenamefont {Greentree}, \citenamefont {Busch},\ and\ \citenamefont
	{Mompart}}]{Menchon2016}%
\BibitemOpen
\bibfield  {author} {\bibinfo {author} {\bibfnamefont {R.}~\bibnamefont
		{Menchon-Enrich}}, \bibinfo {author} {\bibfnamefont {A.}~\bibnamefont
		{Benseny}}, \bibinfo {author} {\bibfnamefont {V.}~\bibnamefont {Ahufinger}},
	\bibinfo {author} {\bibfnamefont {A.~D.}\ \bibnamefont {Greentree}}, \bibinfo
	{author} {\bibfnamefont {Th.}\ \bibnamefont {Busch}}, \ and\ \bibinfo
	{author} {\bibfnamefont {J.}~\bibnamefont {Mompart}},\ }\bibfield  {title}
{\enquote {\bibinfo {title} {{Reports on Progress in Physics Spatial
				adiabatic passage: a review of recent progress Related content}},}\
}\href@noop {} {\bibfield  {journal} {\bibinfo  {journal} {Rep. Prog. Phys.}\
}\textbf {\bibinfo {volume} {79}} (\bibinfo {year} {2016})}\BibitemShut
{NoStop}%
\bibitem [{\citenamefont {Ban}\ \emph {et~al.}(2019)\citenamefont {Ban},
	\citenamefont {Chen}, \citenamefont {Kohler},\ and\ \citenamefont
	{Platero}}]{Platero2019}%
\BibitemOpen
\bibfield  {author} {\bibinfo {author} {\bibfnamefont {Yue}\ \bibnamefont
		{Ban}}, \bibinfo {author} {\bibfnamefont {Xi}~\bibnamefont {Chen}}, \bibinfo
	{author} {\bibfnamefont {Sigmund}\ \bibnamefont {Kohler}}, \ and\ \bibinfo
	{author} {\bibfnamefont {Gloria}\ \bibnamefont {Platero}},\ }\bibfield
{title} {\enquote {\bibinfo {title} {Spin entangled state transfer in quantum
			dot arrays: Coherent adiabatic and speed-up protocols},}\ }\href@noop {}
{\bibfield  {journal} {\bibinfo  {journal} {Advanced Quantum Technologies}\
	}\textbf {\bibinfo {volume} {2}} (\bibinfo {year} {2019})}\BibitemShut
{NoStop}%
\bibitem [{\citenamefont {Petrosyan}\ \emph {et~al.}(2010)\citenamefont
	{Petrosyan}, \citenamefont {Nikolopoulos},\ and\ \citenamefont
	{Lambropoulos}}]{Petrosyan2010}%
\BibitemOpen
\bibfield  {author} {\bibinfo {author} {\bibfnamefont {David}\ \bibnamefont
		{Petrosyan}}, \bibinfo {author} {\bibfnamefont {Georgios~M.}\ \bibnamefont
		{Nikolopoulos}}, \ and\ \bibinfo {author} {\bibfnamefont {P.}~\bibnamefont
		{Lambropoulos}},\ }\bibfield  {title} {\enquote {\bibinfo {title} {State
			transfer in static and dynamic spin chains with disorder},}\ }\href {\doibase
	10.1103/PhysRevA.81.042307} {\bibfield  {journal} {\bibinfo  {journal} {Phys.
			Rev. A}\ }\textbf {\bibinfo {volume} {81}},\ \bibinfo {pages} {042307}
	(\bibinfo {year} {2010})}\BibitemShut {NoStop}%
\bibitem [{\citenamefont {Chancellor}\ and\ \citenamefont
	{Haas}(2012)}]{Chancellor2012}%
\BibitemOpen
\bibfield  {author} {\bibinfo {author} {\bibfnamefont {Nicholas}\
		\bibnamefont {Chancellor}}\ and\ \bibinfo {author} {\bibfnamefont {Stephan}\
		\bibnamefont {Haas}},\ }\bibfield  {title} {\enquote {\bibinfo {title} {Using
			{theJ}1{\textendash}j2quantum spin chain as an adiabatic quantum data bus},}\
}\href {\doibase 10.1088/1367-2630/14/9/095025} {\bibfield  {journal}
{\bibinfo  {journal} {New Journal of Physics}\ }\textbf {\bibinfo {volume}
	{14}},\ \bibinfo {pages} {095025} (\bibinfo {year} {2012})}\BibitemShut
{NoStop}%
\bibitem [{\citenamefont {Farooq}\ \emph {et~al.}(2015)\citenamefont {Farooq},
	\citenamefont {Bayat}, \citenamefont {Mancini},\ and\ \citenamefont
	{Bose}}]{Farooq2015}%
\BibitemOpen
\bibfield  {author} {\bibinfo {author} {\bibfnamefont {Umer}\ \bibnamefont
		{Farooq}}, \bibinfo {author} {\bibfnamefont {Abolfazl}\ \bibnamefont
		{Bayat}}, \bibinfo {author} {\bibfnamefont {Stefano}\ \bibnamefont
		{Mancini}}, \ and\ \bibinfo {author} {\bibfnamefont {Sougato}\ \bibnamefont
		{Bose}},\ }\bibfield  {title} {\enquote {\bibinfo {title} {Adiabatic
			many-body state preparation and information transfer in quantum dot
			arrays},}\ }\href {\doibase 10.1103/PhysRevB.91.134303} {\bibfield  {journal}
	{\bibinfo  {journal} {Phys. Rev. B}\ }\textbf {\bibinfo {volume} {91}},\
	\bibinfo {pages} {134303} (\bibinfo {year} {2015})}\BibitemShut {NoStop}%
\bibitem [{\citenamefont {Vitanov}\ \emph {et~al.}(2017)\citenamefont
	{Vitanov}, \citenamefont {Rangelov}, \citenamefont {Shore},\ and\
	\citenamefont {Bergmann}}]{Vitanov2017}%
\BibitemOpen
\bibfield  {author} {\bibinfo {author} {\bibfnamefont {Nikolay~V.}\
		\bibnamefont {Vitanov}}, \bibinfo {author} {\bibfnamefont {Andon~A.}\
		\bibnamefont {Rangelov}}, \bibinfo {author} {\bibfnamefont {Bruce~W.}\
		\bibnamefont {Shore}}, \ and\ \bibinfo {author} {\bibfnamefont {Klaas}\
		\bibnamefont {Bergmann}},\ }\bibfield  {title} {\enquote {\bibinfo {title}
		{Stimulated raman adiabatic passage in physics, chemistry, and beyond},}\
}\href@noop {} {\bibfield  {journal} {\bibinfo  {journal} {Rev. Mod. Phys.}\
}\textbf {\bibinfo {volume} {89}},\ \bibinfo {pages} {015006} (\bibinfo
{year} {2017})}\BibitemShut {NoStop}%
\bibitem [{\citenamefont {Kumar}\ \emph {et~al.}(2016)\citenamefont {Kumar},
	\citenamefont {Veps{\"a}l{\"a}inen}, \citenamefont {Danilin},\ and\
	\citenamefont {Paraoanu}}]{Kumar2016apsc}%
\BibitemOpen
\bibfield  {author} {\bibinfo {author} {\bibfnamefont {K.~S.}\ \bibnamefont
		{Kumar}}, \bibinfo {author} {\bibfnamefont {A.}~\bibnamefont
		{Veps{\"a}l{\"a}inen}}, \bibinfo {author} {\bibfnamefont {S.}~\bibnamefont
		{Danilin}}, \ and\ \bibinfo {author} {\bibfnamefont {G.~S.}\ \bibnamefont
		{Paraoanu}},\ }\bibfield  {title} {\enquote {\bibinfo {title} {Stimulated
			raman adiabatic passage in a three-level superconducting circuit},}\ }\href
{\doibase 10.1038/ncomms10628} {\bibfield  {journal} {\bibinfo  {journal}
		{Nature Communications}\ }\textbf {\bibinfo {volume} {7}},\ \bibinfo {pages}
	{10628} (\bibinfo {year} {2016})}\BibitemShut {NoStop}%
\bibitem [{\citenamefont {Wunderlich}\ \emph {et~al.}(2007)\citenamefont
	{Wunderlich}, \citenamefont {Hannemann}, \citenamefont {K{\"o}rber},
	\citenamefont {H{\"a}ffner}, \citenamefont {Roos}, \citenamefont
	{H{\"a}nsel}, \citenamefont {Blatt},\ and\ \citenamefont
	{Schmidt-Kaler}}]{wunderlich2007robust}%
\BibitemOpen
\bibfield  {author} {\bibinfo {author} {\bibfnamefont {Chr}\ \bibnamefont
		{Wunderlich}}, \bibinfo {author} {\bibfnamefont {Th}~\bibnamefont
		{Hannemann}}, \bibinfo {author} {\bibfnamefont {T}~\bibnamefont
		{K{\"o}rber}}, \bibinfo {author} {\bibfnamefont {H}~\bibnamefont
		{H{\"a}ffner}}, \bibinfo {author} {\bibfnamefont {Ch}~\bibnamefont {Roos}},
	\bibinfo {author} {\bibfnamefont {W}~\bibnamefont {H{\"a}nsel}}, \bibinfo
	{author} {\bibfnamefont {R}~\bibnamefont {Blatt}}, \ and\ \bibinfo {author}
	{\bibfnamefont {F}~\bibnamefont {Schmidt-Kaler}},\ }\bibfield  {title}
{\enquote {\bibinfo {title} {Robust state preparation of a single trapped ion
			by adiabatic passage},}\ }\href@noop {} {\bibfield  {journal} {\bibinfo
		{journal} {Journal of Modern Optics}\ }\textbf {\bibinfo {volume} {54}},\
	\bibinfo {pages} {1541--1549} (\bibinfo {year} {2007})}\BibitemShut {NoStop}%
\bibitem [{\citenamefont {Veps{\"a}l{\"a}inen}\ \emph
	{et~al.}(2019)\citenamefont {Veps{\"a}l{\"a}inen}, \citenamefont {Danilin},\
	and\ \citenamefont {Paraoanu}}]{Vepsalainen2019superad}%
\BibitemOpen
\bibfield  {author} {\bibinfo {author} {\bibfnamefont {Antti}\ \bibnamefont
		{Veps{\"a}l{\"a}inen}}, \bibinfo {author} {\bibfnamefont {Sergey}\
		\bibnamefont {Danilin}}, \ and\ \bibinfo {author} {\bibfnamefont
		{Gheorghe~Sorin}\ \bibnamefont {Paraoanu}},\ }\bibfield  {title} {\enquote
	{\bibinfo {title} {Superadiabatic population transfer in a three-level
			superconducting circuit},}\ }\href {\doibase 10.1126/sciadv.aau5999}
{\bibfield  {journal} {\bibinfo  {journal} {Science Advances}\ }\textbf
	{\bibinfo {volume} {5}} (\bibinfo {year} {2019}),\
	10.1126/sciadv.aau5999}\BibitemShut {NoStop}%
\bibitem [{sup()}]{supmat}%
\BibitemOpen
\href@noop {} {}\bibinfo {note} {See Suppplemental Material for further
	information on experimental procedures, calculations, and
	simulations.}\BibitemShut {Stop}%
\bibitem [{\citenamefont {Pioro-Ladri{\`e}re}\ \emph
	{et~al.}(2008)\citenamefont {Pioro-Ladri{\`e}re}, \citenamefont {Obata},
	\citenamefont {Tokura}, \citenamefont {Shin}, \citenamefont {Kubo},
	\citenamefont {Yoshida}, \citenamefont {Taniyama},\ and\ \citenamefont
	{Tarucha}}]{Pioro-Ladriere2008}%
\BibitemOpen
\bibfield  {author} {\bibinfo {author} {\bibfnamefont {M.}~\bibnamefont
		{Pioro-Ladri{\`e}re}}, \bibinfo {author} {\bibfnamefont {T.}~\bibnamefont
		{Obata}}, \bibinfo {author} {\bibfnamefont {Y.}~\bibnamefont {Tokura}},
	\bibinfo {author} {\bibfnamefont {Y.-S.}\ \bibnamefont {Shin}}, \bibinfo
	{author} {\bibfnamefont {T.}~\bibnamefont {Kubo}}, \bibinfo {author}
	{\bibfnamefont {K.}~\bibnamefont {Yoshida}}, \bibinfo {author} {\bibfnamefont
		{T.}~\bibnamefont {Taniyama}}, \ and\ \bibinfo {author} {\bibfnamefont
		{S.}~\bibnamefont {Tarucha}},\ }\bibfield  {title} {\enquote {\bibinfo
		{title} {Electrically driven single-electron spin resonance in a slanting
			zeeman field},}\ }\href@noop {} {\bibfield  {journal} {\bibinfo  {journal}
		{Nature Physics}\ }\textbf {\bibinfo {volume} {4}} (\bibinfo {year}
	{2008})}\BibitemShut {NoStop}%
\bibitem [{\citenamefont {Koppens}\ \emph {et~al.}(2006)\citenamefont
	{Koppens}, \citenamefont {Buizert}, \citenamefont {Tielrooij}, \citenamefont
	{Vink}, \citenamefont {Nowack}, \citenamefont {Meunier}, \citenamefont
	{Kouwenhoven},\ and\ \citenamefont {Vandersypen}}]{Koppens2006}%
\BibitemOpen
\bibfield  {author} {\bibinfo {author} {\bibfnamefont {F.~H.~L.}\
		\bibnamefont {Koppens}}, \bibinfo {author} {\bibfnamefont {C.}~\bibnamefont
		{Buizert}}, \bibinfo {author} {\bibfnamefont {K.~J.}\ \bibnamefont
		{Tielrooij}}, \bibinfo {author} {\bibfnamefont {I.~T.}\ \bibnamefont {Vink}},
	\bibinfo {author} {\bibfnamefont {K.~C.}\ \bibnamefont {Nowack}}, \bibinfo
	{author} {\bibfnamefont {T.}~\bibnamefont {Meunier}}, \bibinfo {author}
	{\bibfnamefont {L.~P.}\ \bibnamefont {Kouwenhoven}}, \ and\ \bibinfo {author}
	{\bibfnamefont {L.~M.~K.}\ \bibnamefont {Vandersypen}},\ }\bibfield  {title}
{\enquote {\bibinfo {title} {Driven coherent oscillations of a single
			electron spin in a quantum dot},}\ }\href@noop {} {\bibfield  {journal}
	{\bibinfo  {journal} {Nature}\ }\textbf {\bibinfo {volume} {442}},\ \bibinfo
	{pages} {766--771} (\bibinfo {year} {2006})}\BibitemShut {NoStop}%
\bibitem [{\citenamefont {Gullans}\ and\ \citenamefont
	{Petta}(2020)}]{gullans2020SCTAP}%
\BibitemOpen
\bibfield  {author} {\bibinfo {author} {\bibfnamefont {M.~J.}\ \bibnamefont
		{Gullans}}\ and\ \bibinfo {author} {\bibfnamefont {J.~R.}\ \bibnamefont
		{Petta}},\ }\bibfield  {title} {\enquote {\bibinfo {title} {Coherent
			transport of spin by adiabatic passage in quantum dot arrays},}\ }\href
{\doibase 10.1103/PhysRevB.102.155404} {\bibfield  {journal} {\bibinfo
		{journal} {Phys. Rev. B}\ }\textbf {\bibinfo {volume} {102}},\ \bibinfo
	{pages} {155404} (\bibinfo {year} {2020})}\BibitemShut {NoStop}%
\bibitem [{\citenamefont {Angus}\ \emph {et~al.}(2007)\citenamefont {Angus},
	\citenamefont {Ferguson}, \citenamefont {Dzurak},\ and\ \citenamefont
	{Clark}}]{Angus2007}%
\BibitemOpen
\bibfield  {author} {\bibinfo {author} {\bibfnamefont {Susan~J.}\
		\bibnamefont {Angus}}, \bibinfo {author} {\bibfnamefont {Andrew~J.}\
		\bibnamefont {Ferguson}}, \bibinfo {author} {\bibfnamefont {Andrew~S.}\
		\bibnamefont {Dzurak}}, \ and\ \bibinfo {author} {\bibfnamefont {Robert~G.}\
		\bibnamefont {Clark}},\ }\bibfield  {title} {\enquote {\bibinfo {title}
		{Gate-defined quantum dots in intrinsic silicon},}\ }\href {\doibase
	10.1021/nl070949k} {\bibfield  {journal} {\bibinfo  {journal} {Nano Letters}\
	}\textbf {\bibinfo {volume} {7}},\ \bibinfo {pages} {2051--2055} (\bibinfo
	{year} {2007})}\BibitemShut {NoStop}%
\bibitem [{\citenamefont {Barthel}\ \emph {et~al.}(2010)\citenamefont
	{Barthel}, \citenamefont {Kj\ae{}rgaard}, \citenamefont {Medford},
	\citenamefont {Stopa}, \citenamefont {Marcus}, \citenamefont {Hanson},\ and\
	\citenamefont {Gossard}}]{Barthel2010}%
\BibitemOpen
\bibfield  {author} {\bibinfo {author} {\bibfnamefont {C.}~\bibnamefont
		{Barthel}}, \bibinfo {author} {\bibfnamefont {M.}~\bibnamefont
		{Kj\ae{}rgaard}}, \bibinfo {author} {\bibfnamefont {J.}~\bibnamefont
		{Medford}}, \bibinfo {author} {\bibfnamefont {M.}~\bibnamefont {Stopa}},
	\bibinfo {author} {\bibfnamefont {C.~M.}\ \bibnamefont {Marcus}}, \bibinfo
	{author} {\bibfnamefont {M.~P.}\ \bibnamefont {Hanson}}, \ and\ \bibinfo
	{author} {\bibfnamefont {A.~C.}\ \bibnamefont {Gossard}},\ }\bibfield
{title} {\enquote {\bibinfo {title} {Fast sensing of double-dot charge
			arrangement and spin state with a radio-frequency sensor quantum dot},}\
}\href@noop {} {\bibfield  {journal} {\bibinfo  {journal} {Phys. Rev. B}\
}\textbf {\bibinfo {volume} {81}},\ \bibinfo {pages} {161308} (\bibinfo
{year} {2010})}\BibitemShut {NoStop}%
\bibitem [{\citenamefont {Studenikin}\ \emph {et~al.}(2012)\citenamefont
	{Studenikin}, \citenamefont {Thorgrimson}, \citenamefont {Aers},
	\citenamefont {Kam}, \citenamefont {Zawadzki}, \citenamefont {Wasilewski},
	\citenamefont {Bogan},\ and\ \citenamefont {Sachrajda}}]{Studenikin2012}%
\BibitemOpen
\bibfield  {author} {\bibinfo {author} {\bibfnamefont {S.~A.}\ \bibnamefont
		{Studenikin}}, \bibinfo {author} {\bibfnamefont {J.}~\bibnamefont
		{Thorgrimson}}, \bibinfo {author} {\bibfnamefont {G.~C.}\ \bibnamefont
		{Aers}}, \bibinfo {author} {\bibfnamefont {A.}~\bibnamefont {Kam}}, \bibinfo
	{author} {\bibfnamefont {P.}~\bibnamefont {Zawadzki}}, \bibinfo {author}
	{\bibfnamefont {Z.~R.}\ \bibnamefont {Wasilewski}}, \bibinfo {author}
	{\bibfnamefont {A.}~\bibnamefont {Bogan}}, \ and\ \bibinfo {author}
	{\bibfnamefont {A.~S.}\ \bibnamefont {Sachrajda}},\ }\bibfield  {title}
{\enquote {\bibinfo {title} {Enhanced charge detection of spin qubit readout
			via an intermediate state},}\ }\href {\doibase 10.1063/1.4749281} {\bibfield
	{journal} {\bibinfo  {journal} {Applied Physics Letters}\ }\textbf {\bibinfo
		{volume} {101}},\ \bibinfo {pages} {233101} (\bibinfo {year}
	{2012})}\BibitemShut {NoStop}%
\bibitem [{\citenamefont {Reed}\ \emph {et~al.}(2016)\citenamefont {Reed},
	\citenamefont {Maune}, \citenamefont {Andrews}, \citenamefont {Borselli},
	\citenamefont {Eng}, \citenamefont {Jura}, \citenamefont {Kiselev},
	\citenamefont {Ladd}, \citenamefont {Merkel}, \citenamefont {Milosavljevic},
	\citenamefont {Pritchett}, \citenamefont {Rakher}, \citenamefont {Ross},
	\citenamefont {Schmitz}, \citenamefont {Smith}, \citenamefont {Wright},
	\citenamefont {Gyure},\ and\ \citenamefont {Hunter}}]{Reed2016SymSi}%
\BibitemOpen
\bibfield  {author} {\bibinfo {author} {\bibfnamefont {M.~D.}\ \bibnamefont
		{Reed}}, \bibinfo {author} {\bibfnamefont {B.~M.}\ \bibnamefont {Maune}},
	\bibinfo {author} {\bibfnamefont {R.~W.}\ \bibnamefont {Andrews}}, \bibinfo
	{author} {\bibfnamefont {M.~G.}\ \bibnamefont {Borselli}}, \bibinfo {author}
	{\bibfnamefont {K.}~\bibnamefont {Eng}}, \bibinfo {author} {\bibfnamefont
		{M.~P.}\ \bibnamefont {Jura}}, \bibinfo {author} {\bibfnamefont {A.~A.}\
		\bibnamefont {Kiselev}}, \bibinfo {author} {\bibfnamefont {T.~D.}\
		\bibnamefont {Ladd}}, \bibinfo {author} {\bibfnamefont {S.~T.}\ \bibnamefont
		{Merkel}}, \bibinfo {author} {\bibfnamefont {I.}~\bibnamefont
		{Milosavljevic}}, \bibinfo {author} {\bibfnamefont {E.~J.}\ \bibnamefont
		{Pritchett}}, \bibinfo {author} {\bibfnamefont {M.~T.}\ \bibnamefont
		{Rakher}}, \bibinfo {author} {\bibfnamefont {R.~S.}\ \bibnamefont {Ross}},
	\bibinfo {author} {\bibfnamefont {A.~E.}\ \bibnamefont {Schmitz}}, \bibinfo
	{author} {\bibfnamefont {A.}~\bibnamefont {Smith}}, \bibinfo {author}
	{\bibfnamefont {J.~A.}\ \bibnamefont {Wright}}, \bibinfo {author}
	{\bibfnamefont {M.~F.}\ \bibnamefont {Gyure}}, \ and\ \bibinfo {author}
	{\bibfnamefont {A.~T.}\ \bibnamefont {Hunter}},\ }\bibfield  {title}
{\enquote {\bibinfo {title} {Reduced sensitivity to charge noise in
			semiconductor spin qubits via symmetric operation},}\ }\href {\doibase
	10.1103/PhysRevLett.116.110402} {\bibfield  {journal} {\bibinfo  {journal}
		{Phys. Rev. Lett.}\ }\textbf {\bibinfo {volume} {116}},\ \bibinfo {pages}
	{110402} (\bibinfo {year} {2016})}\BibitemShut {NoStop}%
\bibitem [{\citenamefont {Martins}\ \emph {et~al.}(2016)\citenamefont
	{Martins}, \citenamefont {Malinowski}, \citenamefont {Nissen}, \citenamefont
	{Barnes}, \citenamefont {Fallahi}, \citenamefont {Gardner}, \citenamefont
	{Manfra}, \citenamefont {Marcus},\ and\ \citenamefont
	{Kuemmeth}}]{Martins2016}%
\BibitemOpen
\bibfield  {author} {\bibinfo {author} {\bibfnamefont {Frederico}\
		\bibnamefont {Martins}}, \bibinfo {author} {\bibfnamefont {Filip~K.}\
		\bibnamefont {Malinowski}}, \bibinfo {author} {\bibfnamefont {Peter~D.}\
		\bibnamefont {Nissen}}, \bibinfo {author} {\bibfnamefont {Edwin}\
		\bibnamefont {Barnes}}, \bibinfo {author} {\bibfnamefont {Saeed}\
		\bibnamefont {Fallahi}}, \bibinfo {author} {\bibfnamefont {Geoffrey~C.}\
		\bibnamefont {Gardner}}, \bibinfo {author} {\bibfnamefont {Michael~J.}\
		\bibnamefont {Manfra}}, \bibinfo {author} {\bibfnamefont {Charles~M.}\
		\bibnamefont {Marcus}}, \ and\ \bibinfo {author} {\bibfnamefont {Ferdinand}\
		\bibnamefont {Kuemmeth}},\ }\bibfield  {title} {\enquote {\bibinfo {title}
		{Noise suppression using symmetric exchange gates in spin qubits},}\ }\href
{\doibase 10.1103/PhysRevLett.116.116801} {\bibfield  {journal} {\bibinfo
		{journal} {Phys. Rev. Lett.}\ }\textbf {\bibinfo {volume} {116}},\ \bibinfo
	{pages} {116801} (\bibinfo {year} {2016})}\BibitemShut {NoStop}%
\bibitem [{\citenamefont {Petta}\ \emph {et~al.}(2005)\citenamefont {Petta},
	\citenamefont {Johnson}, \citenamefont {Taylor}, \citenamefont {Laird},
	\citenamefont {Yacoby}, \citenamefont {Lukin}, \citenamefont {Marcus},
	\citenamefont {Hanson},\ and\ \citenamefont {Gossard}}]{Petta2005}%
\BibitemOpen
\bibfield  {author} {\bibinfo {author} {\bibfnamefont {J.~R.}\ \bibnamefont
		{Petta}}, \bibinfo {author} {\bibfnamefont {A.~C.}\ \bibnamefont {Johnson}},
	\bibinfo {author} {\bibfnamefont {J.~M.}\ \bibnamefont {Taylor}}, \bibinfo
	{author} {\bibfnamefont {E.~A.}\ \bibnamefont {Laird}}, \bibinfo {author}
	{\bibfnamefont {A.}~\bibnamefont {Yacoby}}, \bibinfo {author} {\bibfnamefont
		{M.~D.}\ \bibnamefont {Lukin}}, \bibinfo {author} {\bibfnamefont {C.~M.}\
		\bibnamefont {Marcus}}, \bibinfo {author} {\bibfnamefont {M.~P.}\
		\bibnamefont {Hanson}}, \ and\ \bibinfo {author} {\bibfnamefont {A.~C.}\
		\bibnamefont {Gossard}},\ }\bibfield  {title} {\enquote {\bibinfo {title}
		{Coherent manipulation of coupled electron spins in semiconductor quantum
			dots},}\ }\href {\doibase 10.1126/science.1116955} {\bibfield  {journal}
	{\bibinfo  {journal} {Science}\ }\textbf {\bibinfo {volume} {309}},\ \bibinfo
	{pages} {2180--2184} (\bibinfo {year} {2005})}\BibitemShut {NoStop}%
\bibitem [{\citenamefont {Foletti}\ \emph
	{et~al.}(2009{\natexlab{a}})\citenamefont {Foletti}, \citenamefont {Bluhm},
	\citenamefont {Mahalu}, \citenamefont {Umansky},\ and\ \citenamefont
	{Yacoby}}]{foletti2009universal}%
\BibitemOpen
\bibfield  {author} {\bibinfo {author} {\bibfnamefont {Sandra}\ \bibnamefont
		{Foletti}}, \bibinfo {author} {\bibfnamefont {Hendrik}\ \bibnamefont
		{Bluhm}}, \bibinfo {author} {\bibfnamefont {Diana}\ \bibnamefont {Mahalu}},
	\bibinfo {author} {\bibfnamefont {Vladimir}\ \bibnamefont {Umansky}}, \ and\
	\bibinfo {author} {\bibfnamefont {Amir}\ \bibnamefont {Yacoby}},\ }\bibfield
{title} {\enquote {\bibinfo {title} {Universal quantum control of
			two-electron spin quantum bits using dynamic nuclear polarization},}\
}\href@noop {} {\bibfield  {journal} {\bibinfo  {journal} {Nature Physics}\
}\textbf {\bibinfo {volume} {5}},\ \bibinfo {pages} {903--908} (\bibinfo
{year} {2009}{\natexlab{a}})}\BibitemShut {NoStop}%
\bibitem [{\citenamefont {Reilly}\ \emph {et~al.}(2010)\citenamefont {Reilly},
	\citenamefont {Taylor}, \citenamefont {Petta}, \citenamefont {Marcus},
	\citenamefont {Hanson},\ and\ \citenamefont {Gossard}}]{Reilly2010}%
\BibitemOpen
\bibfield  {author} {\bibinfo {author} {\bibfnamefont {D.~J.}\ \bibnamefont
		{Reilly}}, \bibinfo {author} {\bibfnamefont {J.~M.}\ \bibnamefont {Taylor}},
	\bibinfo {author} {\bibfnamefont {J.~R.}\ \bibnamefont {Petta}}, \bibinfo
	{author} {\bibfnamefont {C.~M.}\ \bibnamefont {Marcus}}, \bibinfo {author}
	{\bibfnamefont {M.~P.}\ \bibnamefont {Hanson}}, \ and\ \bibinfo {author}
	{\bibfnamefont {A.~C.}\ \bibnamefont {Gossard}},\ }\bibfield  {title}
{\enquote {\bibinfo {title} {Exchange control of nuclear spin diffusion in a
			double quantum dot},}\ }\href {\doibase 10.1103/PhysRevLett.104.236802}
{\bibfield  {journal} {\bibinfo  {journal} {Phys. Rev. Lett.}\ }\textbf
	{\bibinfo {volume} {104}},\ \bibinfo {pages} {236802} (\bibinfo {year}
	{2010})}\BibitemShut {NoStop}%
\bibitem [{\citenamefont {Shulman}\ \emph {et~al.}(2014)\citenamefont
	{Shulman}, \citenamefont {Harvey}, \citenamefont {Nichol}, \citenamefont
	{Bartlett}, \citenamefont {Doherty}, \citenamefont {Umansky},\ and\
	\citenamefont {Yacoby}}]{Shulman2014}%
\BibitemOpen
\bibfield  {author} {\bibinfo {author} {\bibfnamefont {M.~D.}\ \bibnamefont
		{Shulman}}, \bibinfo {author} {\bibfnamefont {S.~P.}\ \bibnamefont {Harvey}},
	\bibinfo {author} {\bibfnamefont {J.~M.}\ \bibnamefont {Nichol}}, \bibinfo
	{author} {\bibfnamefont {S.~D.}\ \bibnamefont {Bartlett}}, \bibinfo {author}
	{\bibfnamefont {A.~C.}\ \bibnamefont {Doherty}}, \bibinfo {author}
	{\bibfnamefont {V.}~\bibnamefont {Umansky}}, \ and\ \bibinfo {author}
	{\bibfnamefont {A.}~\bibnamefont {Yacoby}},\ }\bibfield  {title} {\enquote
	{\bibinfo {title} {Suppressing qubit dephasing using real-time hamiltonian
			estimation},}\ }\href {\doibase 10.1038/ncomms6156} {\bibfield  {journal}
	{\bibinfo  {journal} {Nature Communications}\ }\textbf {\bibinfo {volume}
		{5}},\ \bibinfo {pages} {5156} (\bibinfo {year} {2014})}\BibitemShut
{NoStop}%
\bibitem [{\citenamefont {Takeda}\ \emph {et~al.}(2020)\citenamefont {Takeda},
	\citenamefont {Noiri}, \citenamefont {Yoneda}, \citenamefont {Nakajima},\
	and\ \citenamefont {Tarucha}}]{Takeda2020STsi}%
\BibitemOpen
\bibfield  {author} {\bibinfo {author} {\bibfnamefont {K.}~\bibnamefont
		{Takeda}}, \bibinfo {author} {\bibfnamefont {A.}~\bibnamefont {Noiri}},
	\bibinfo {author} {\bibfnamefont {J.}~\bibnamefont {Yoneda}}, \bibinfo
	{author} {\bibfnamefont {T.}~\bibnamefont {Nakajima}}, \ and\ \bibinfo
	{author} {\bibfnamefont {S.}~\bibnamefont {Tarucha}},\ }\bibfield  {title}
{\enquote {\bibinfo {title} {Resonantly driven singlet-triplet spin qubit in
			silicon},}\ }\href {\doibase 10.1103/PhysRevLett.124.117701} {\bibfield
	{journal} {\bibinfo  {journal} {Phys. Rev. Lett.}\ }\textbf {\bibinfo
		{volume} {124}},\ \bibinfo {pages} {117701} (\bibinfo {year}
	{2020})}\BibitemShut {NoStop}%
\bibitem [{\citenamefont {Jones}\ \emph {et~al.}(2019)\citenamefont {Jones},
	\citenamefont {Pritchett}, \citenamefont {Chen}, \citenamefont {Keating},
	\citenamefont {Andrews}, \citenamefont {Blumoff}, \citenamefont {De~Lorenzo},
	\citenamefont {Eng}, \citenamefont {Ha}, \citenamefont {Kiselev},
	\citenamefont {Meenehan}, \citenamefont {Merkel}, \citenamefont {Wright},
	\citenamefont {Edge}, \citenamefont {Ross}, \citenamefont {Rakher},
	\citenamefont {Borselli},\ and\ \citenamefont {Hunter}}]{Jones2019}%
\BibitemOpen
\bibfield  {author} {\bibinfo {author} {\bibfnamefont {A.M.}\ \bibnamefont
		{Jones}}, \bibinfo {author} {\bibfnamefont {E.J.}\ \bibnamefont {Pritchett}},
	\bibinfo {author} {\bibfnamefont {E.H.}\ \bibnamefont {Chen}}, \bibinfo
	{author} {\bibfnamefont {T.E.}\ \bibnamefont {Keating}}, \bibinfo {author}
	{\bibfnamefont {R.W.}\ \bibnamefont {Andrews}}, \bibinfo {author}
	{\bibfnamefont {J.Z.}\ \bibnamefont {Blumoff}}, \bibinfo {author}
	{\bibfnamefont {L.A.}\ \bibnamefont {De~Lorenzo}}, \bibinfo {author}
	{\bibfnamefont {K.}~\bibnamefont {Eng}}, \bibinfo {author} {\bibfnamefont
		{S.D.}\ \bibnamefont {Ha}}, \bibinfo {author} {\bibfnamefont {A.A.}\
		\bibnamefont {Kiselev}}, \bibinfo {author} {\bibfnamefont {S.M.}\
		\bibnamefont {Meenehan}}, \bibinfo {author} {\bibfnamefont {S.T.}\
		\bibnamefont {Merkel}}, \bibinfo {author} {\bibfnamefont {J.A.}\ \bibnamefont
		{Wright}}, \bibinfo {author} {\bibfnamefont {L.F.}\ \bibnamefont {Edge}},
	\bibinfo {author} {\bibfnamefont {R.S.}\ \bibnamefont {Ross}}, \bibinfo
	{author} {\bibfnamefont {M.T.}\ \bibnamefont {Rakher}}, \bibinfo {author}
	{\bibfnamefont {M.G.}\ \bibnamefont {Borselli}}, \ and\ \bibinfo {author}
	{\bibfnamefont {A.}~\bibnamefont {Hunter}},\ }\bibfield  {title} {\enquote
	{\bibinfo {title} {Spin-blockade spectroscopy of
			$\mathrm{Si}$/$\mathrm{Si}$-$\mathrm{Ge}$ quantum dots},}\ }\href {\doibase
	10.1103/PhysRevApplied.12.014026} {\bibfield  {journal} {\bibinfo  {journal}
		{Phys. Rev. Applied}\ }\textbf {\bibinfo {volume} {12}},\ \bibinfo {pages}
	{014026} (\bibinfo {year} {2019})}\BibitemShut {NoStop}%
\bibitem [{\citenamefont {Connors}\ \emph {et~al.}(2020)\citenamefont
	{Connors}, \citenamefont {Nelson},\ and\ \citenamefont
	{Nichol}}]{Elliot2020}%
\BibitemOpen
\bibfield  {author} {\bibinfo {author} {\bibfnamefont {Elliot~J.}\
		\bibnamefont {Connors}}, \bibinfo {author} {\bibfnamefont {JJ}~\bibnamefont
		{Nelson}}, \ and\ \bibinfo {author} {\bibfnamefont {John~M.}\ \bibnamefont
		{Nichol}},\ }\bibfield  {title} {\enquote {\bibinfo {title} {Rapid
			high-fidelity spin-state readout in $\mathrm{Si}$/$\mathrm{Si}$-$\mathrm{Ge}$
			quantum dots via rf reflectometry},}\ }\href@noop {} {\bibfield  {journal}
	{\bibinfo  {journal} {Phys. Rev. Applied}\ }\textbf {\bibinfo {volume}
		{13}},\ \bibinfo {pages} {024019} (\bibinfo {year} {2020})}\BibitemShut
{NoStop}%
\bibitem [{\citenamefont {Massar}\ and\ \citenamefont
	{Popescu}(1995)}]{Massar1995}%
\BibitemOpen
\bibfield  {author} {\bibinfo {author} {\bibfnamefont {S.}~\bibnamefont
		{Massar}}\ and\ \bibinfo {author} {\bibfnamefont {S.}~\bibnamefont
		{Popescu}},\ }\bibfield  {title} {\enquote {\bibinfo {title} {Optimal
			extraction of information from finite quantum ensembles},}\ }\href {\doibase
	10.1103/PhysRevLett.74.1259} {\bibfield  {journal} {\bibinfo  {journal}
		{Phys. Rev. Lett.}\ }\textbf {\bibinfo {volume} {74}},\ \bibinfo {pages}
	{1259--1263} (\bibinfo {year} {1995})}\BibitemShut {NoStop}%
\bibitem [{\citenamefont {Steffen}\ \emph {et~al.}(2013)\citenamefont
	{Steffen}, \citenamefont {Salathe}, \citenamefont {Oppliger}, \citenamefont
	{Kurpiers}, \citenamefont {Baur}, \citenamefont {Lang}, \citenamefont
	{Eichler}, \citenamefont {Puebla-Hellmann}, \citenamefont {Fedorov},\ and\
	\citenamefont {Wallraff}}]{Steffen2013}%
\BibitemOpen
\bibfield  {author} {\bibinfo {author} {\bibfnamefont {L.}~\bibnamefont
		{Steffen}}, \bibinfo {author} {\bibfnamefont {Y.}~\bibnamefont {Salathe}},
	\bibinfo {author} {\bibfnamefont {M.}~\bibnamefont {Oppliger}}, \bibinfo
	{author} {\bibfnamefont {P.}~\bibnamefont {Kurpiers}}, \bibinfo {author}
	{\bibfnamefont {M.}~\bibnamefont {Baur}}, \bibinfo {author} {\bibfnamefont
		{C.}~\bibnamefont {Lang}}, \bibinfo {author} {\bibfnamefont {C.}~\bibnamefont
		{Eichler}}, \bibinfo {author} {\bibfnamefont {G.}~\bibnamefont
		{Puebla-Hellmann}}, \bibinfo {author} {\bibfnamefont {A.}~\bibnamefont
		{Fedorov}}, \ and\ \bibinfo {author} {\bibfnamefont {A.}~\bibnamefont
		{Wallraff}},\ }\bibfield  {title} {\enquote {\bibinfo {title} {Deterministic
			quantum teleportation with feed-forward in a solid state system},}\ }\href
{\doibase 10.1038/nature12422} {\bibfield  {journal} {\bibinfo  {journal}
		{Nature}\ }\textbf {\bibinfo {volume} {500}},\ \bibinfo {pages} {319--322}
	(\bibinfo {year} {2013})}\BibitemShut {NoStop}%
\bibitem [{\citenamefont {Oh}\ \emph {et~al.}(2010)\citenamefont {Oh},
	\citenamefont {Friesen},\ and\ \citenamefont {Hu}}]{Oh2010}%
\BibitemOpen
\bibfield  {author} {\bibinfo {author} {\bibfnamefont {Sangchul}\
		\bibnamefont {Oh}}, \bibinfo {author} {\bibfnamefont {Mark}\ \bibnamefont
		{Friesen}}, \ and\ \bibinfo {author} {\bibfnamefont {Xuedong}\ \bibnamefont
		{Hu}},\ }\bibfield  {title} {\enquote {\bibinfo {title} {Even-odd effects of
			heisenberg chains on long-range interaction and entanglement},}\ }\href
{\doibase 10.1103/PhysRevB.82.140403} {\bibfield  {journal} {\bibinfo
		{journal} {Phys. Rev. B}\ }\textbf {\bibinfo {volume} {82}},\ \bibinfo
	{pages} {140403} (\bibinfo {year} {2010})}\BibitemShut {NoStop}%
\bibitem [{\citenamefont {Oh}\ \emph {et~al.}(2011)\citenamefont {Oh},
	\citenamefont {Wu}, \citenamefont {Shim}, \citenamefont {Fei}, \citenamefont
	{Friesen},\ and\ \citenamefont {Hu}}]{Oh2011}%
\BibitemOpen
\bibfield  {author} {\bibinfo {author} {\bibfnamefont {Sangchul}\
		\bibnamefont {Oh}}, \bibinfo {author} {\bibfnamefont {Lian-Ao}\ \bibnamefont
		{Wu}}, \bibinfo {author} {\bibfnamefont {Yun-Pil}\ \bibnamefont {Shim}},
	\bibinfo {author} {\bibfnamefont {Jianjia}\ \bibnamefont {Fei}}, \bibinfo
	{author} {\bibfnamefont {Mark}\ \bibnamefont {Friesen}}, \ and\ \bibinfo
	{author} {\bibfnamefont {Xuedong}\ \bibnamefont {Hu}},\ }\bibfield  {title}
{\enquote {\bibinfo {title} {Heisenberg spin bus as a robust transmission
			line for quantum-state transfer},}\ }\href {\doibase
	10.1103/PhysRevA.84.022330} {\bibfield  {journal} {\bibinfo  {journal} {Phys.
			Rev. A}\ }\textbf {\bibinfo {volume} {84}},\ \bibinfo {pages} {022330}
	(\bibinfo {year} {2011})}\BibitemShut {NoStop}%
\bibitem [{\citenamefont {de~Sousa}\ \emph {et~al.}(2001)\citenamefont
	{de~Sousa}, \citenamefont {Hu},\ and\ \citenamefont {Das~Sarma}}]{Sousa2001}%
\BibitemOpen
\bibfield  {author} {\bibinfo {author} {\bibfnamefont {Rogerio}\ \bibnamefont
		{de~Sousa}}, \bibinfo {author} {\bibfnamefont {Xuedong}\ \bibnamefont {Hu}},
	\ and\ \bibinfo {author} {\bibfnamefont {S.}~\bibnamefont {Das~Sarma}},\
}\bibfield  {title} {\enquote {\bibinfo {title} {Effect of an inhomogeneous
		external magnetic field on a quantum-dot quantum computer},}\ }\href
{\doibase 10.1103/PhysRevA.64.042307} {\bibfield  {journal} {\bibinfo
		{journal} {Phys. Rev. A}\ }\textbf {\bibinfo {volume} {64}},\ \bibinfo
	{pages} {042307} (\bibinfo {year} {2001})}\BibitemShut {NoStop}%
\bibitem [{\citenamefont {Foletti}\ \emph
	{et~al.}(2009{\natexlab{b}})\citenamefont {Foletti}, \citenamefont {Bluhm},
	\citenamefont {Mahalu}, \citenamefont {Umansky},\ and\ \citenamefont
	{Yacoby}}]{Foletti2009}%
\BibitemOpen
\bibfield  {author} {\bibinfo {author} {\bibfnamefont {Sandra}\ \bibnamefont
		{Foletti}}, \bibinfo {author} {\bibfnamefont {Hendrik}\ \bibnamefont
		{Bluhm}}, \bibinfo {author} {\bibfnamefont {Diana}\ \bibnamefont {Mahalu}},
	\bibinfo {author} {\bibfnamefont {Vladimir}\ \bibnamefont {Umansky}}, \ and\
	\bibinfo {author} {\bibfnamefont {Amir}\ \bibnamefont {Yacoby}},\ }\bibfield
{title} {\enquote {\bibinfo {title} {Universal quantum control of
			two-electron spin quantum bits using dynamic nuclear polarization},}\
}\href@noop {} {\bibfield  {journal} {\bibinfo  {journal} {Nature Physics}\
}\textbf {\bibinfo {volume} {5}},\ \bibinfo {pages} {903--908} (\bibinfo
{year} {2009}{\natexlab{b}})}\BibitemShut {NoStop}%
\bibitem [{\citenamefont {Orona}\ \emph {et~al.}(2018)\citenamefont {Orona},
	\citenamefont {Nichol}, \citenamefont {Harvey}, \citenamefont {B\o{}ttcher},
	\citenamefont {Fallahi}, \citenamefont {Gardner}, \citenamefont {Manfra},\
	and\ \citenamefont {Yacoby}}]{Orona2018Tp}%
\BibitemOpen
\bibfield  {author} {\bibinfo {author} {\bibfnamefont {Lucas~A.}\
		\bibnamefont {Orona}}, \bibinfo {author} {\bibfnamefont {John~M.}\
		\bibnamefont {Nichol}}, \bibinfo {author} {\bibfnamefont {Shannon~P.}\
		\bibnamefont {Harvey}}, \bibinfo {author} {\bibfnamefont {Charlotte G.~L.}\
		\bibnamefont {B\o{}ttcher}}, \bibinfo {author} {\bibfnamefont {Saeed}\
		\bibnamefont {Fallahi}}, \bibinfo {author} {\bibfnamefont {Geoffrey~C.}\
		\bibnamefont {Gardner}}, \bibinfo {author} {\bibfnamefont {Michael~J.}\
		\bibnamefont {Manfra}}, \ and\ \bibinfo {author} {\bibfnamefont {Amir}\
		\bibnamefont {Yacoby}},\ }\bibfield  {title} {\enquote {\bibinfo {title}
		{Readout of singlet-triplet qubits at large magnetic field gradients},}\
}\href {\doibase 10.1103/PhysRevB.98.125404} {\bibfield  {journal} {\bibinfo
	{journal} {Phys. Rev. B}\ }\textbf {\bibinfo {volume} {98}},\ \bibinfo
{pages} {125404} (\bibinfo {year} {2018})}\BibitemShut {NoStop}%
\bibitem [{\citenamefont {Jock}\ \emph {et~al.}(2018)\citenamefont {Jock},
	\citenamefont {Jacobson}, \citenamefont {Harvey-Collard}, \citenamefont
	{Mounce}, \citenamefont {Srinivasa}, \citenamefont {Ward}, \citenamefont
	{Anderson}, \citenamefont {Manginell}, \citenamefont {Wendt}, \citenamefont
	{Rudolph}, \citenamefont {Pluym}, \citenamefont {Gamble}, \citenamefont
	{Baczewski}, \citenamefont {Witzel},\ and\ \citenamefont
	{Carroll}}]{Jock2018}%
\BibitemOpen
\bibfield  {author} {\bibinfo {author} {\bibfnamefont {Ryan~M.}\ \bibnamefont
		{Jock}}, \bibinfo {author} {\bibfnamefont {N.~Tobias}\ \bibnamefont
		{Jacobson}}, \bibinfo {author} {\bibfnamefont {Patrick}\ \bibnamefont
		{Harvey-Collard}}, \bibinfo {author} {\bibfnamefont {Andrew~M.}\ \bibnamefont
		{Mounce}}, \bibinfo {author} {\bibfnamefont {Vanita}\ \bibnamefont
		{Srinivasa}}, \bibinfo {author} {\bibfnamefont {Dan~R.}\ \bibnamefont
		{Ward}}, \bibinfo {author} {\bibfnamefont {John}\ \bibnamefont {Anderson}},
	\bibinfo {author} {\bibfnamefont {Ron}\ \bibnamefont {Manginell}}, \bibinfo
	{author} {\bibfnamefont {Joel~R.}\ \bibnamefont {Wendt}}, \bibinfo {author}
	{\bibfnamefont {Martin}\ \bibnamefont {Rudolph}}, \bibinfo {author}
	{\bibfnamefont {Tammy}\ \bibnamefont {Pluym}}, \bibinfo {author}
	{\bibfnamefont {John~King}\ \bibnamefont {Gamble}}, \bibinfo {author}
	{\bibfnamefont {Andrew~D.}\ \bibnamefont {Baczewski}}, \bibinfo {author}
	{\bibfnamefont {Wayne~M.}\ \bibnamefont {Witzel}}, \ and\ \bibinfo {author}
	{\bibfnamefont {Malcolm~S.}\ \bibnamefont {Carroll}},\ }\bibfield  {title}
{\enquote {\bibinfo {title} {A silicon metal-oxide-semiconductor electron
			spin-orbit qubit},}\ }\href@noop {} {\bibfield  {journal} {\bibinfo
		{journal} {Nature Communications}\ }\textbf {\bibinfo {volume} {9}},\
	\bibinfo {pages} {1768} (\bibinfo {year} {2018})}\BibitemShut {NoStop}%
\bibitem [{\citenamefont {Chuang}\ and\ \citenamefont
	{Nielsen}(1997)}]{chuang1997}%
\BibitemOpen
\bibfield  {author} {\bibinfo {author} {\bibfnamefont {Isaac~L.}\
		\bibnamefont {Chuang}}\ and\ \bibinfo {author} {\bibfnamefont {M.~A.}\
		\bibnamefont {Nielsen}},\ }\bibfield  {title} {\enquote {\bibinfo {title}
		{Prescription for experimental determination of the dynamics of a quantum
			black box},}\ }\href {\doibase 10.1080/09500349708231894} {\bibfield
	{journal} {\bibinfo  {journal} {Journal of Modern Optics}\ }\textbf {\bibinfo
		{volume} {44}},\ \bibinfo {pages} {2455--2467} (\bibinfo {year}
	{1997})}\BibitemShut {NoStop}%
\end{thebibliography}
\end{document}


\renewcommand{\figurename}{Supplementary Figure}
\renewcommand{\thefigure}{\arabic{figure}}
\renewcommand{\tablename}{Supplementary Table}

\title{Supplementary Information for\\ Adiabatic quantum state transfer in a semiconductor quantum-dot spin chain}

\author{Yadav P. Kandel}
%
\author{Haifeng Qiao}
%
\affiliation{Department of Physics and Astronomy, University of Rochester, Rochester, NY, 14627 USA}
\author{Saeed Fallahi}
\affiliation{Department of Physics and Astronomy, Purdue University, West Lafayette, IN, 47907 USA}
\affiliation{Birck Nanotechnology Center, Purdue University, West Lafayette, IN, 47907 USA}
\author{Geoffrey C. Gardner}
\affiliation{Birck Nanotechnology Center, Purdue University, West Lafayette, IN, 47907 USA}
\affiliation{School of Materials Engineering, Purdue University, West Lafayette, IN, 47907 USA}
\author{Michael J. Manfra}
\affiliation{Department of Physics and Astronomy, Purdue University, West Lafayette, IN, 47907 USA}
\affiliation{Birck Nanotechnology Center, Purdue University, West Lafayette, IN, 47907 USA}
\affiliation{School of Materials Engineering, Purdue University, West Lafayette, IN, 47907 USA}
\affiliation{School of Electrical and Computer Engineering, Purdue University, West Lafayette, IN, 47907 USA}
\author{John M. Nichol}
%
\affiliation{Department of Physics and Astronomy, University of Rochester, Rochester, NY, 14627 USA}
\affiliation{Corresponding author: john.nichol@rochester.edu}


\maketitle

\begin{table*}
	\centering
	\begin{tabular}{|c|c|c|}
		\hline
		Figure in the main text & Hyperfine field (MHz) & Hyperfine field noise (MHz)\\
		\hline
		3(c)&[6.50, -8.00, 60.00, -30.00] & [0.10, 0.10, 0.10, 0.10] \\
		\hline
		3(d)&[36.70, 10.10, -22.10, 16.20] & [0.10, 0.10, 0.10, 0.10] \\
		\hline
		4(c)&[-17.54, -16.12, -1.84, 31.92] & [0.10, 0.10, 0.10, 0.10] \\
		\hline
		4(d)&[-10.52, -17.55, 3.65, 24.11] & [0.10, 0.10, 0.10, 0.10] \\
		\hline
		5(c)-(d)&[ 22.00, 0.00, 20.00, 70.00] & [10.00, 10.00, 10.00, 10.00] \\
		\hline
	\end{tabular}
	\caption{\label{table1} Specific values of the hyperfine field and noise at each dot location used in different simulations.}
\end{table*}
\begin{figure*}
	\includegraphics{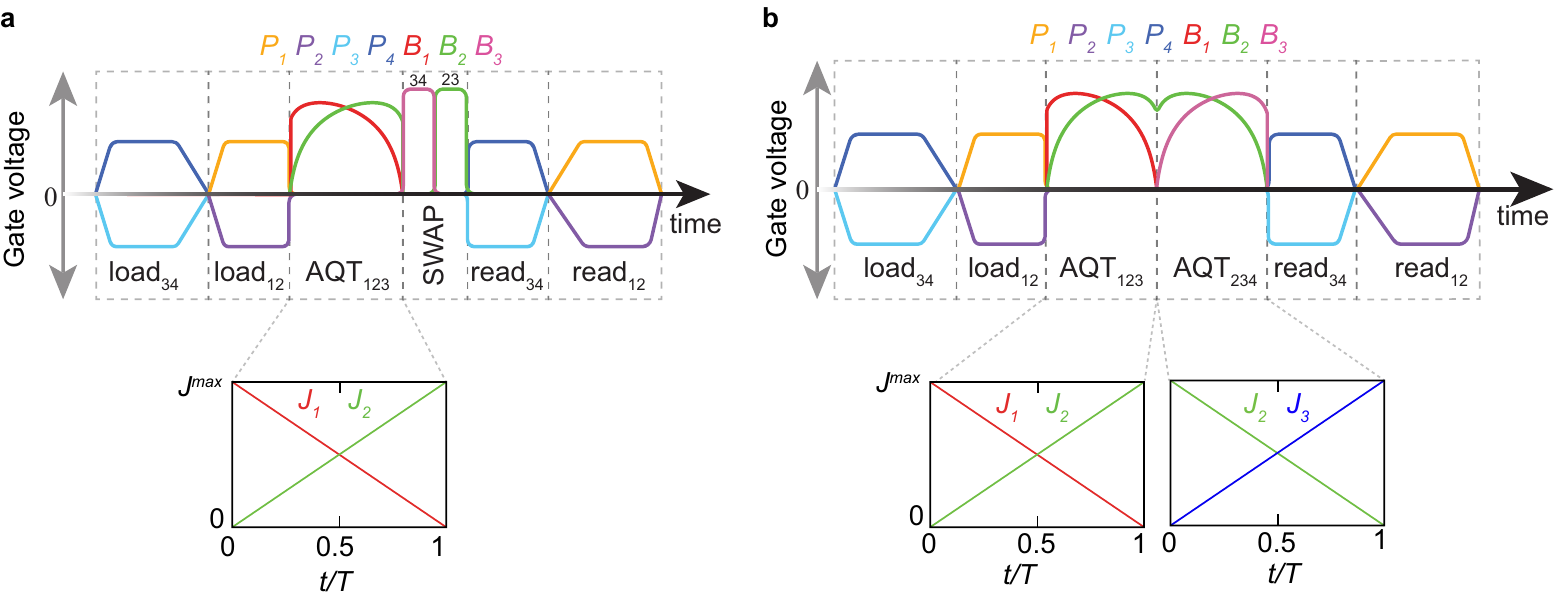}
	\caption{Schematic of pulse diagrams for the AQT experiments. (a) Pulse timing diagram corresponding to Fig. 3(a) in the main text. (b) Pulse timing diagram corresponding to the Fig. 4(a) in the main text. Pulses for the modified method to project a two-electron state onto the singlet/triplet ($ST$) basis are omitted for clarity.}
	\label{orthogonalControl}
\end{figure*}

\begin{figure*}
	\includegraphics{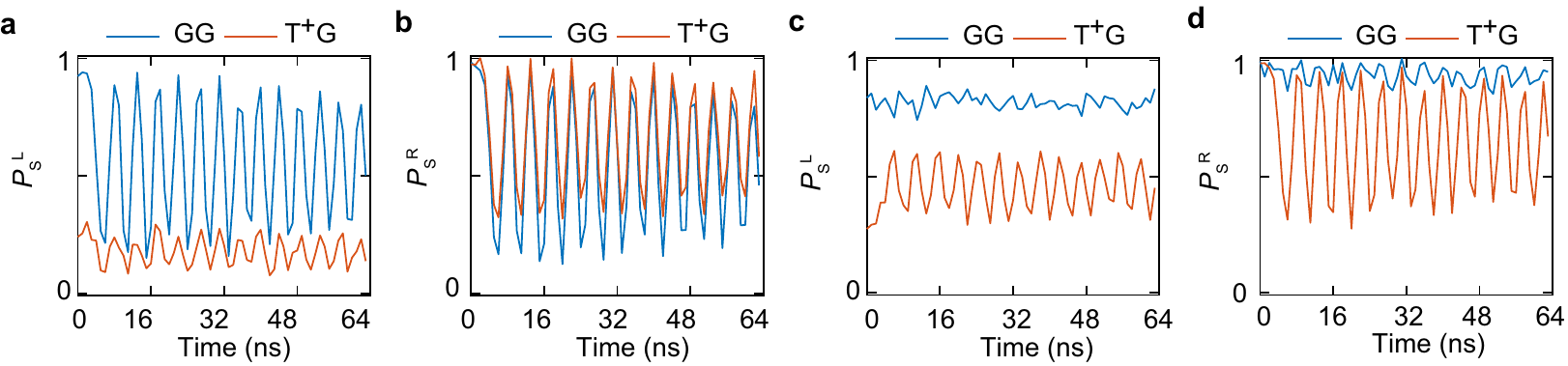}
	\caption{Exchange oscillations of spins 2-3 with different load conditions measured on (a, c) the left side  and (b, d) the right side. The measurements of panels (a-b) were interleaved together and averaged for 256 single shot measurements. The measurements of panels (c-d) were also interleaved. In panels (a-b), prominent oscillations in the case of an $GG$ initialization imply $f=+1$. Prominent oscillations in $P_S^R$ associated with the $T^+G$ load prove that the ground-state spin orientation in dots 3-4 is $\ket{\D\U}$. From this, we can also infer that the ground state spin configuration in dots 1-2 is $\ket{\D\U}$. Panels (c-d) show a similar data set to panels (a-b), but these data were taken at a different time with a different hyperfine configuration. The oscillations in $P_S^R$ associated with the $T^+G$ initialization, and the absence of oscillations associated with the $GG$ initialization, imply that the ground-state spin configuration of the spin chain is $\ket{\U\D\D\U}$, and $f=-1$.} 
	\label{spinGSTATEsup}
\end{figure*}

\begin{figure*}
\includegraphics{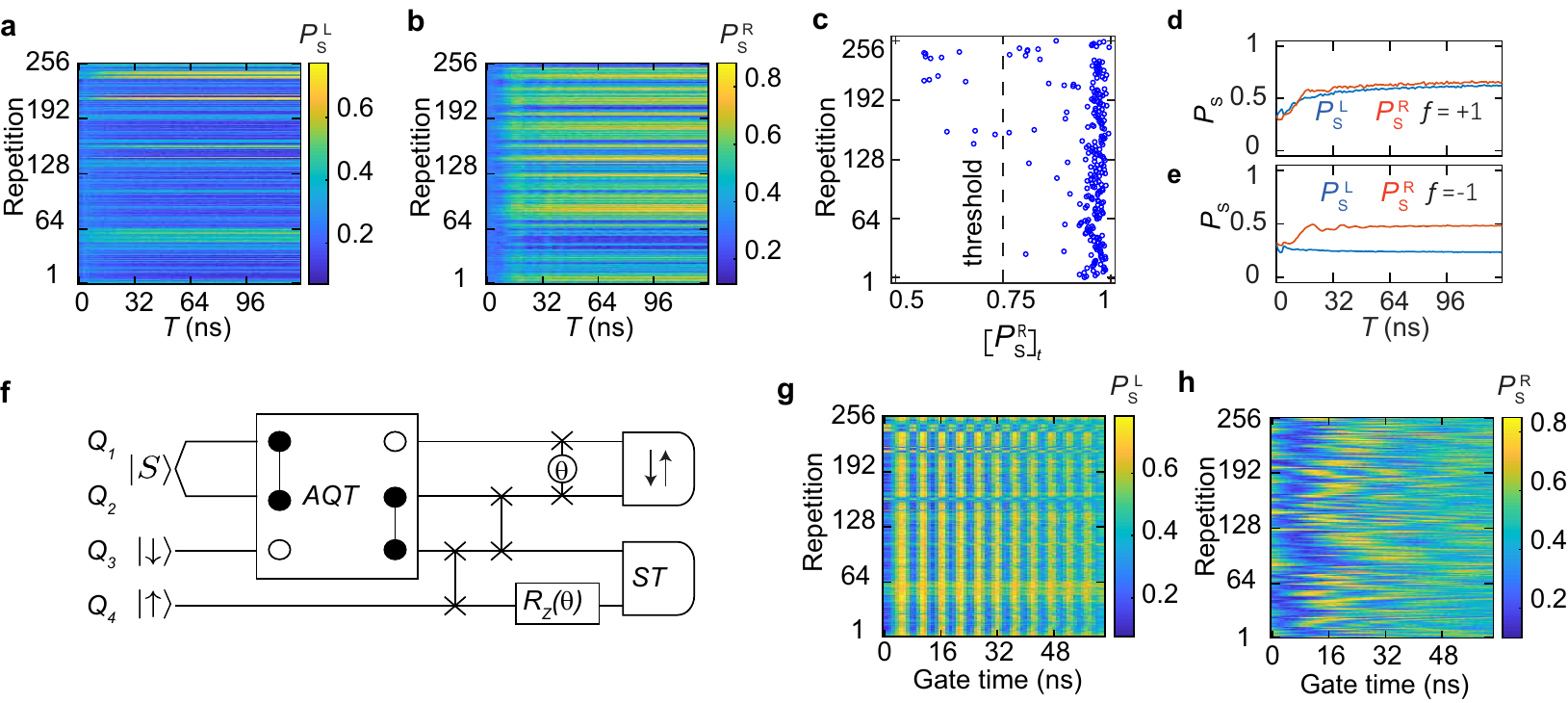}
\caption{(a-b) Data for all repetitions of the experiment described in Fig.~3 in the main text. The line-to-line variations are caused by hyperfine fluctuations, which affect the AQT and the SWAP gates. (c) Time-averaged right-side singlet return probabilities $\ta{P_S^R}$ associated with the evolution of spins 2-3 under the exchange coupling after $GG$ initialization. These measurements were used to determine $f$. (d-e) Averages over all repetitions of the data in (a-b) corresponding to $f=\pm1$. (f) Circuit diagram for the verification experiments. The spins are initialized in the state $\ket{S_{12}\U_3\D_4}$ or $\ket{S_{12}\D_3\U_4}$. After the AQT and SWAP operations, spins 1-2 evolve under exchange and spins $3-4$ evolve under $\Delta B_{34}^z$ for variable amounts of time. (g-h) Results of the verification experiments. The exchange oscillations in (g) and the singlet-triplet oscillations in (h) provide further evidence of the success of the AQT. In (g), the phase of the oscillations is opposite for $f=\pm1$, as expected. The  frequency of the singlet-triplet oscillations in (h) changes between repetitions because of the fluctuating $\Delta B_{34}^z$. The main experiment, the verification experiment, and measurements to monitor the spin ground were interleaved in time. In panels (a), (b), (g), and (h), each line is averaged over 512 single shot measurements.}
\label{AQTandSWAPsup}
\end{figure*}

\begin{figure*}
\includegraphics{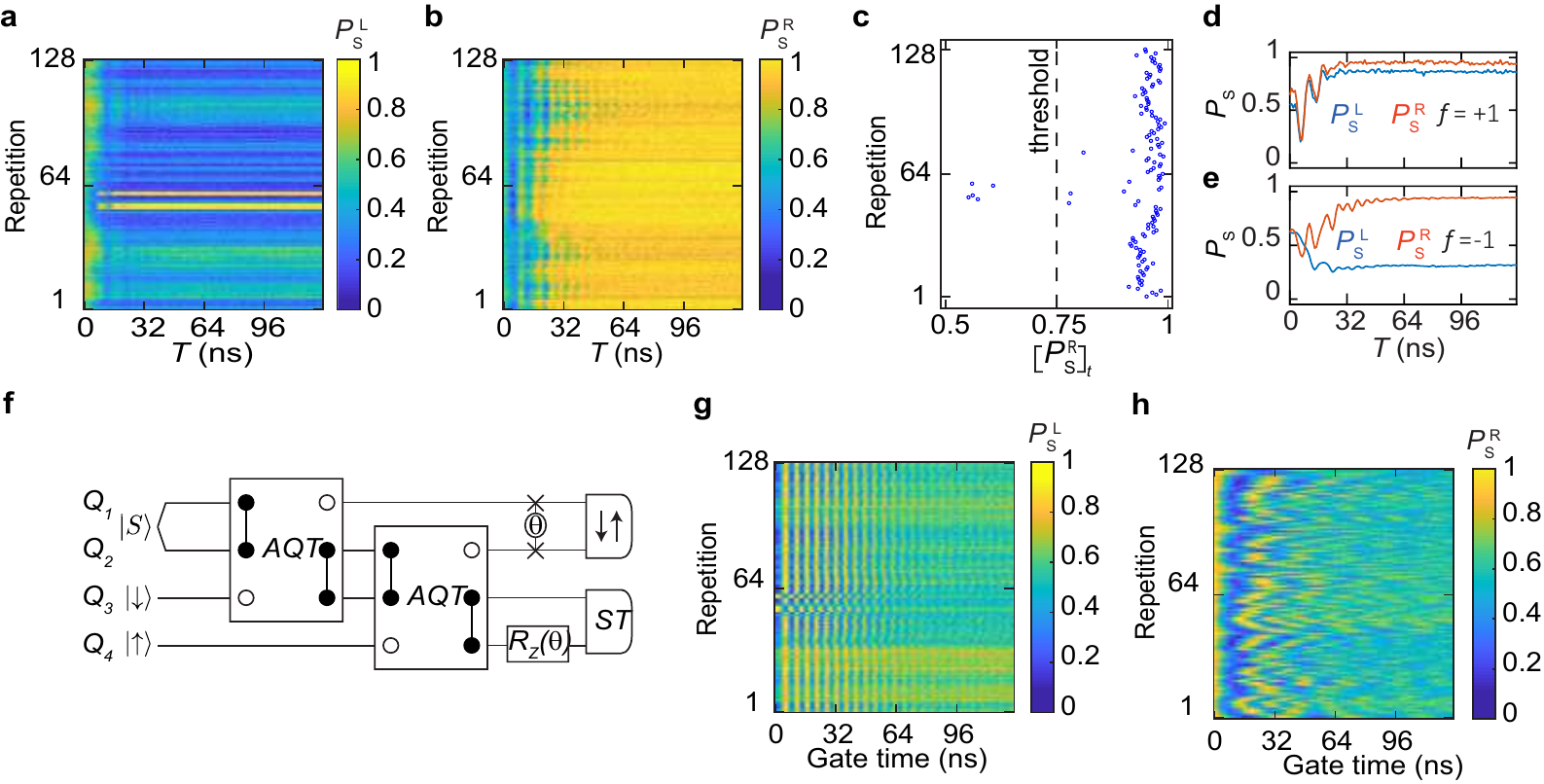}
\caption{(a-b) Data for all repetitions of the AQT cascade experiment described in Fig.~4 in the main text. (c) Time-averaged right-side singlet return probabilities $\ta{P_S^R}$ associated with the evolution of spins 2-3 under the exchange coupling after $GG$ initialization. These measurements were used to determine $f$. (d-e) Averages over all repetitions of the data in (a-b) corresponding to $f=\pm1$. (f) Circuit diagram for the verification experiments. The spins are initialized in the state $\ket{S_{12}\U_3\D_4}$ or $\ket{S_{12}\D_3\U_4}$. After the AQT cascade, spins 1-2 evolve under exchange and spins 3-4 evolve under $\Delta B_{34}^z$ for variable amounts of time. (g-h) Results of the verification experiments. The exchange oscillations in (g) and the singlet-triplet oscillations in (h) provide further evidence of the success of the AQT. In (g), the phase of the oscillations is opposite for $f=\pm1$, as expected. The  frequency of the singlet-triplet oscillations in (h) changes between repetitions because of the fluctuating $\Delta B_{34}^z$. The main experiment, the verification experiment, and measurements to monitor the spin ground were interleaved in time. In panels (a), (b), (g), and (h), each line is averaged over 256 single shot measurements.}
\label{cascadedAQTsup}
\end{figure*}

\begin{figure*}
\includegraphics{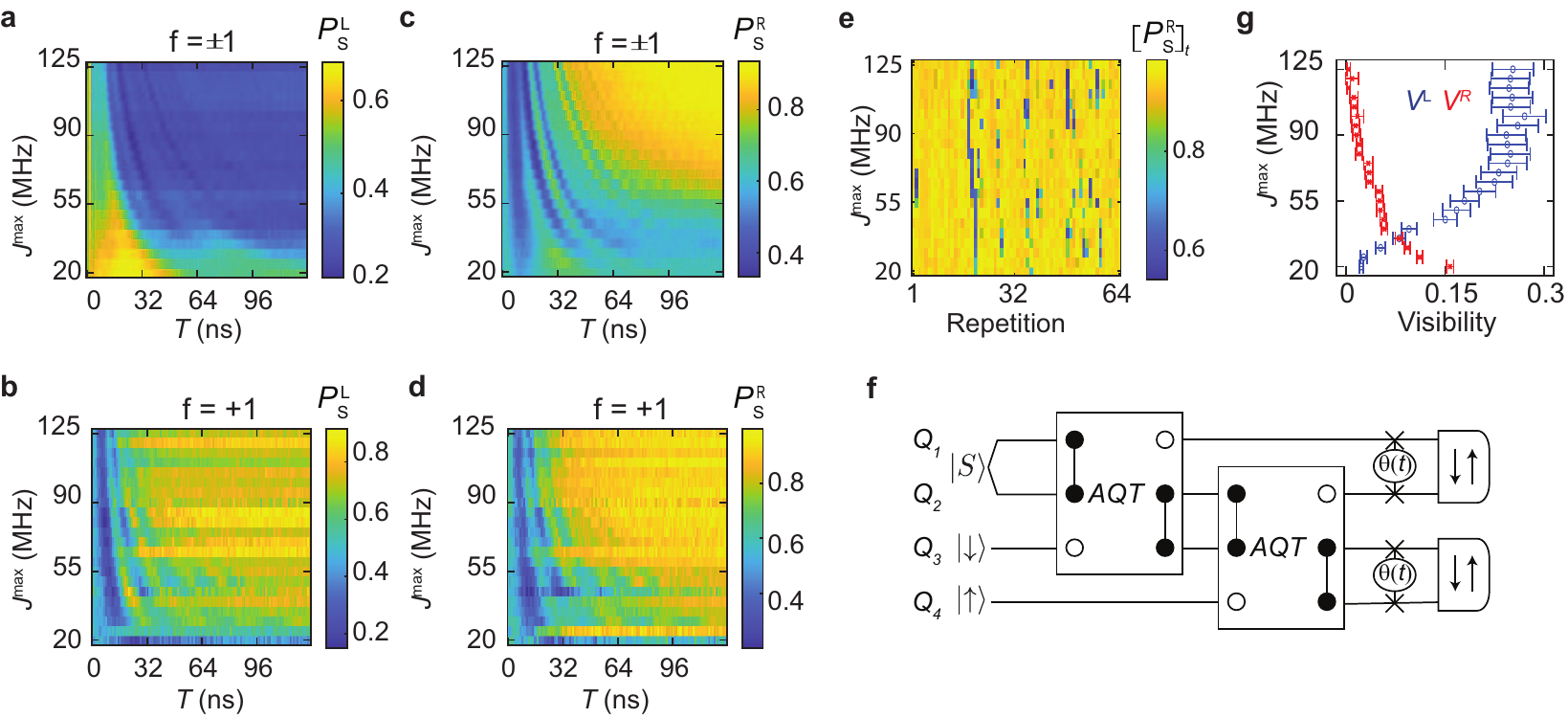}
\caption{\label{2D4SAQTsup} Data for the experiment described in Fig.~5 in the main text. (a,c) Raw data for the left and the right sides. Each line is the average of 128 single shot measurements which are repeated 64 times. Out of all those 64 repetitions for 22 line corresponding to 22 different values of the $J^{max}$, ~95$\%$ correspond to the $f=-1$ which is presented in the Fig.5 in the main text.  (b,d) Post-selected data for the left and the right sides for $f = +1$. (e) Time-averaged right-side singlet return probability $\ta{P_S^R}$ associated with the evolution of spins 2 and 3 under exchange coupling corresponding to $GG$ initialization. We bin the main data set into $f=\pm1$ cases by thresholding the data according to the measurements of panel (e), as discussed above. In most of the cases $f$ is $-1$. (f) Quantum circuit diagram to monitor exchange oscillation visibility after the AQT cascade. (g) Exchange oscillation visibility $V^{R/L}$ of the right and the left spins pairs for different $J^{max}$ with $T=127$~ns for $f=- 1$. $V^L$ increases gradually with $J^{max}$ and saturates. }

\end{figure*}


\begin{figure*}
\includegraphics{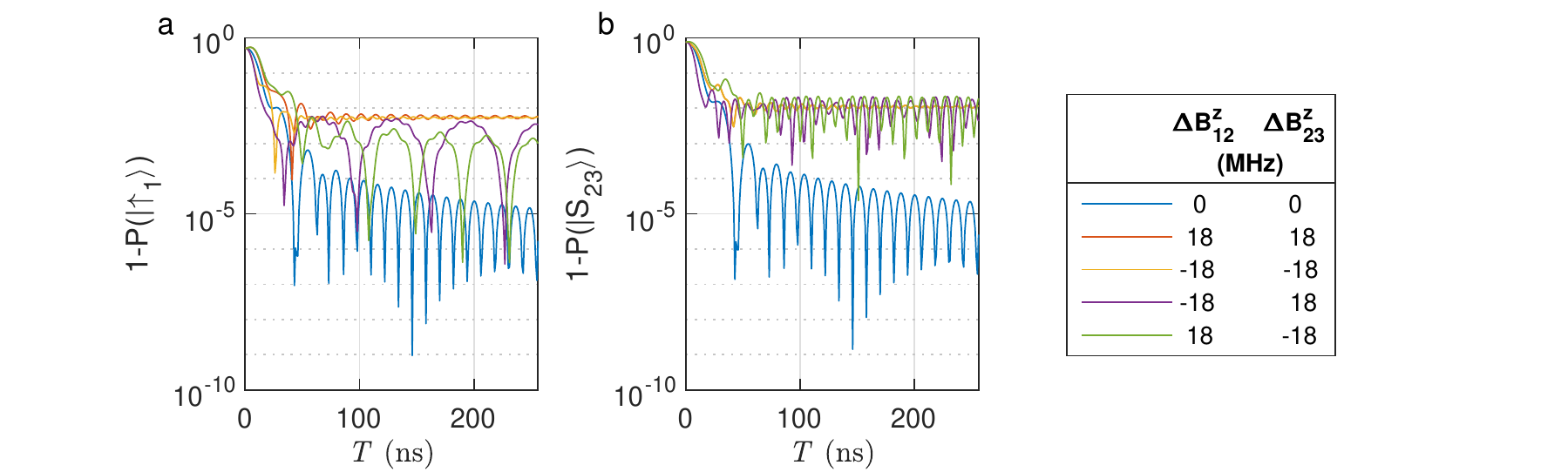}
\caption{\label{supFig7} Simulated effects of a static magnetic field gradient on AQT fidelity. We simulated the evolution of three-spin chain initialized in the state $\ket{S_{12}\U_3}$. We ramped down the exchange coupling between spins 1 and 2 from 120 MHz to 0 while the exchange coupling between spins 2 and 3 was ramped up from 0 to 120 MHz in a time $T$. (a) Infidelity of  transferring the eigenstate of spin 3 to spin 1 as a function of $T$. (b) Infidelity of transferring the singlet state from spin 1-2 to 2-3. No magnetic or charge noise was included in these simulations. The simulations show that the AQT fidelity depends sensitively on $\Delta B^z$.}
\end{figure*}

\begin{figure*}
	\includegraphics{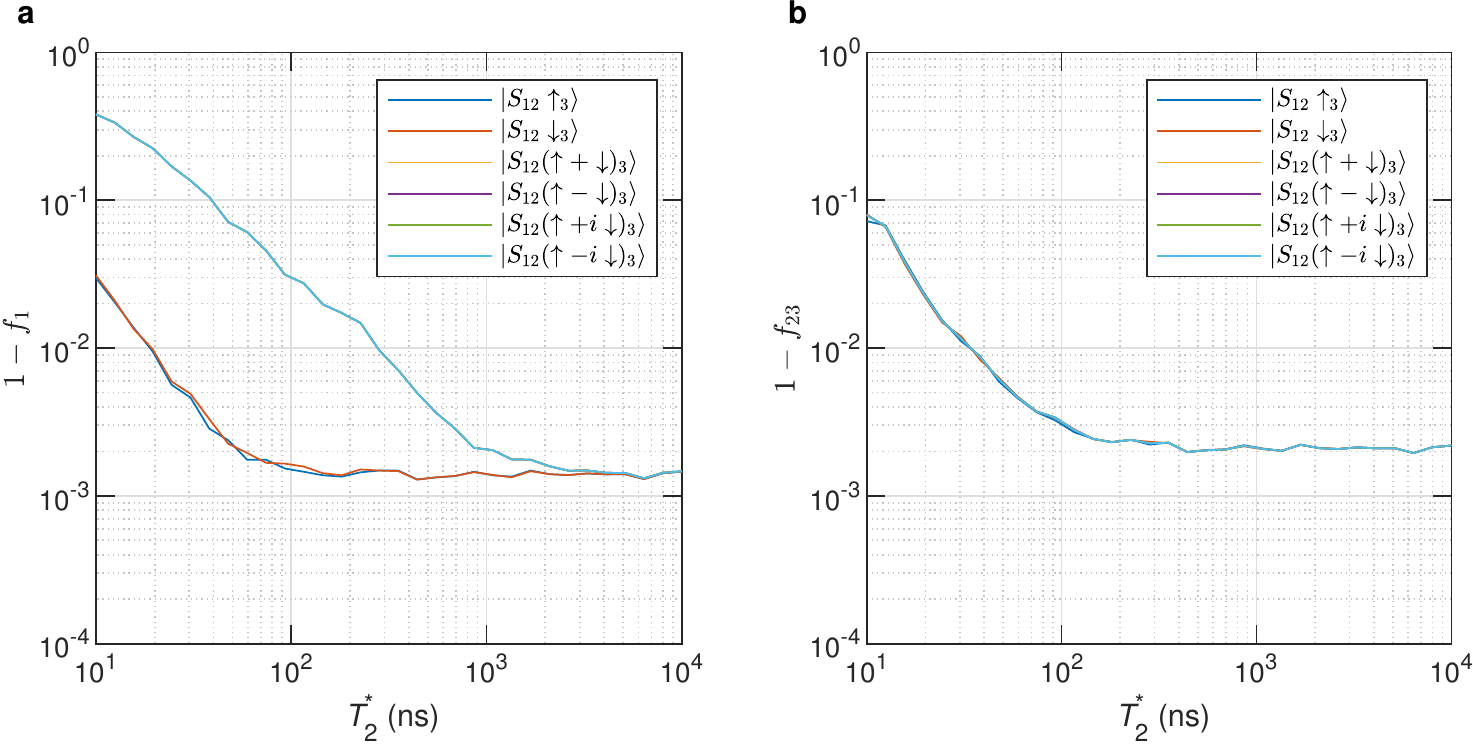}
	\caption{\label{supFig8} Simulated error in the AQT process vs. single-spin $T_2^*$ values for states of the form $\ket{S_{12}\phi_3}$, where $\ket{\phi}$ can be a superposition state. A typical $T_2^*$ value for GaAs is 10-20 ns. For isotopically purified Si, $T_2^*$ can exceed 1 $\mu$s. (a) Probability to incorrectly transfer a single-spin state from dot 3 to dot 1. All of the initial states with superposition states of spin 3 have an error similar to the blue curve. (b) Probability to incorrectly transfer the state of dots 1-2 to 2-3. For isotopically purified Si, these simulations suggest that the state-transfer probability can exceed 0.99. State normalization factors have been omitted in the legends. }
\end{figure*}

\begin{figure*}
	\includegraphics{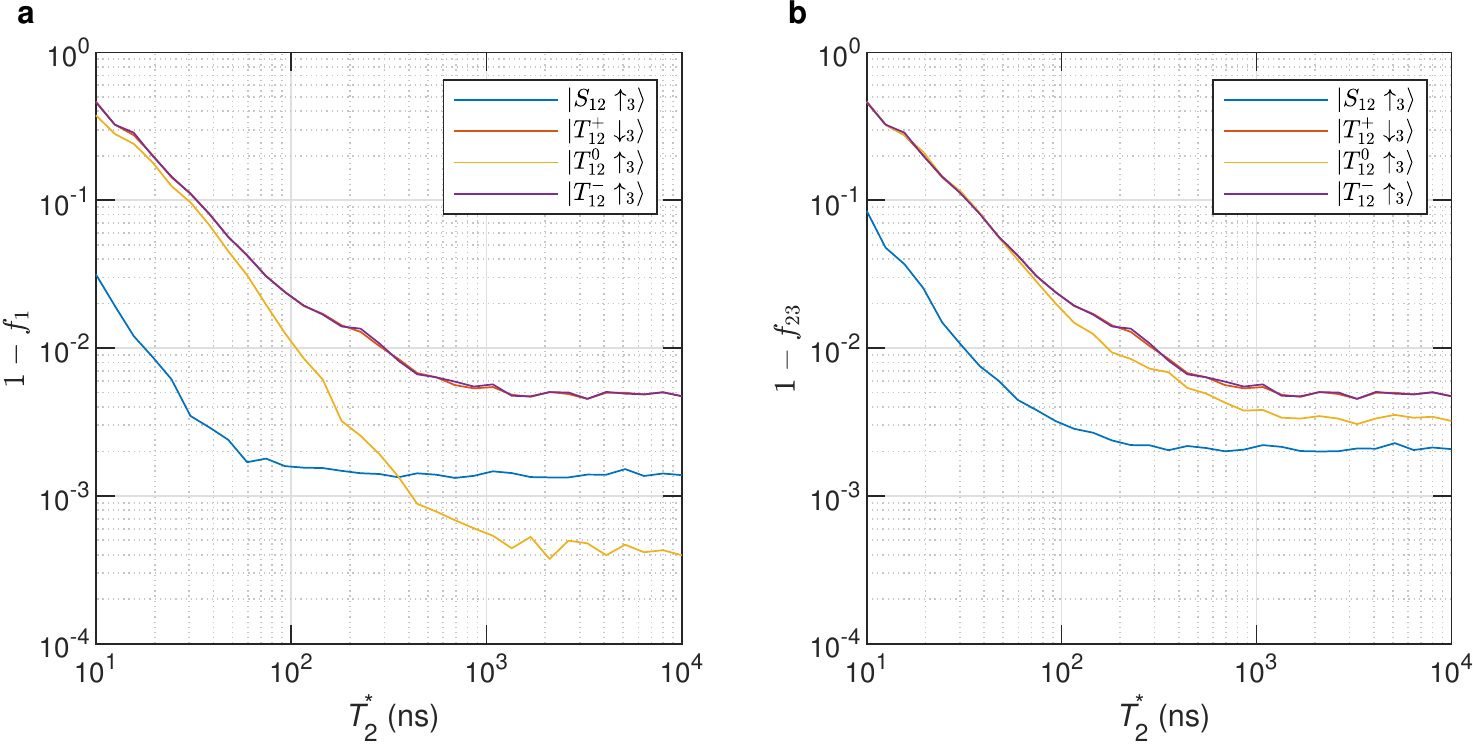}
	\caption{\label{supFig9} Simulated error in the AQT process vs. single-spin $T_2^*$ values for states of the form $\ket{\pi_{12}\phi_3}$, where $\ket{\pi}$ can be a eigenstate of exchange, and $\ket{\phi}$ is a single-spin eigenstate. A typical $T_2^*$ value for GaAs is 10-20 ns. For isotopically purified Si, $T_2^*$ can exceed 1 $\mu$s. (a) Probability to incorrectly transfer a single-spin state from dot 3 to dot 1. (b) Probability to incorrectly transfer the state of dots 1-2 to 2-3. For isotopically purified Si, these simulations suggest that the state-transfer probability can exceed 0.99. }
\end{figure*}

\begin{figure*}
	\includegraphics{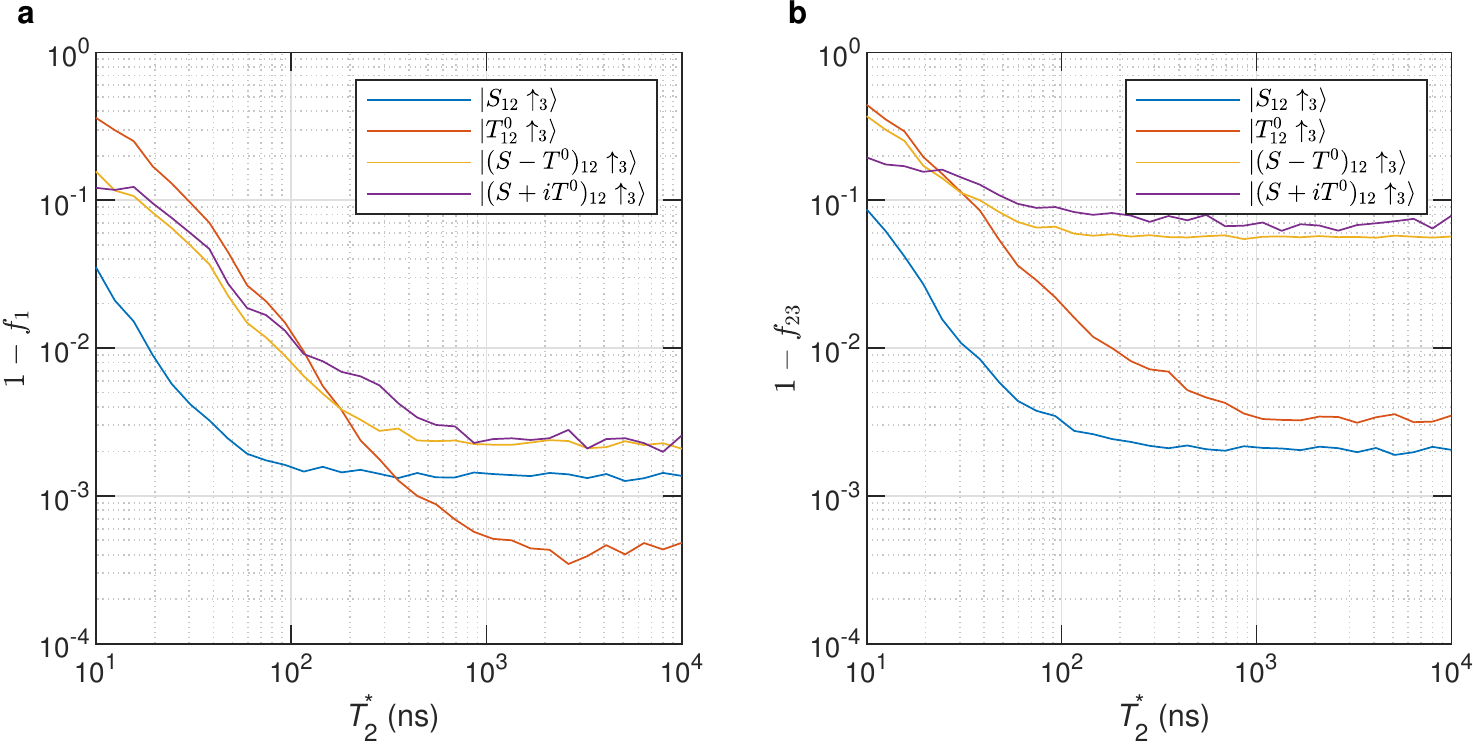}
	\caption{\label{supFig10} Simulated error in the AQT process vs. single-spin $T_2^*$ values for states of the form $\ket{\pi_{12}\U_3}$, where $\ket{\pi}$ can be a superposition of exchange eigenstates. A typical $T_2^*$ value for GaAs is 10-20 ns. For isotopically purified Si, $T_2^*$ can exceed 1 $\mu$s. (a) Probability to incorrectly transfer a single-spin state from dot 3 to dot 1. (b) Probability to incorrectly transfer the state of dots 1-2 to 2-3. For isotopically purified Si, these simulations suggest that the single-spin-eigenstate transfer probability can exceed 0.99. State normalization factors have been omitted in the legends }
\end{figure*}

\begin{figure*}
	\includegraphics{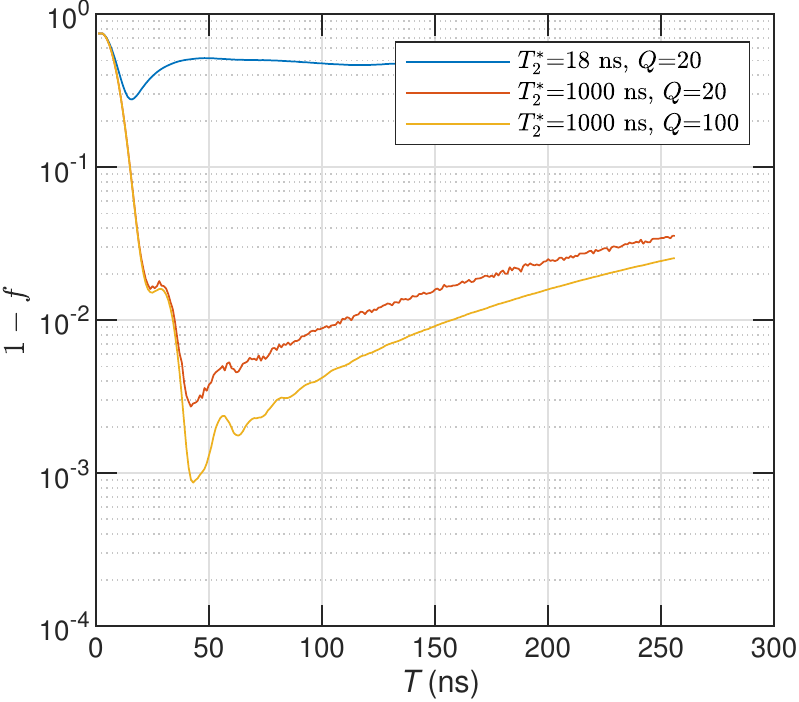}
	\caption{\label{supFigTomo}  Simulated process infidelity for different noise configurations versus Hamiltonian interpolation time $T$. }
\end{figure*}